\documentclass[12pt]{article}

\usepackage[margin=1.3in]{geometry}
\usepackage{graphicx}
\usepackage{subfig}
\usepackage[colorlinks=true,linkcolor=blue,citecolor=red]{hyperref}
\usepackage[square, numbers]{natbib}
\usepackage[ruled]{algorithm2e}
\usepackage{algorithmic}
\usepackage{amsmath}
\usepackage{amssymb}
\usepackage{amsthm}
\usepackage{authblk}
\usepackage{ulem}
\usepackage{soul}

\theoremstyle{definition}

\DeclareMathOperator{\tr}{Tr}

\newcommand{\argmax}[1]{\underset{#1}{\operatorname{arg}\,\operatorname{max}}\;}
\newcommand{\argmin}[1]{\underset{#1}{\operatorname{arg}\,\operatorname{min}}\;}

\newcommand{\norm}[1]{\lVert{#1}\rVert}

\renewcommand{\a}{{\bf a}}
\renewcommand{\b}{{\bf b}}
\renewcommand{\c}{{\bf c}}

\newcommand{\n}{{\bf n}}

\renewcommand{\v}{{\bf v}}

\newcommand{\x}{{\bf x}}
\newcommand{\y}{{\bf y}}
\newcommand{\z}{{\bf z}}

\newcommand{\A}{{\bf A}}
\newcommand{\B}{{\bf B}}
\newcommand{\C}{{\bf R}}

\newcommand{\I}{{\bf I}}

\newcommand{\M}{{\bf M}}
\newcommand{\N}{{\bf N}}

\renewcommand{\P}{{\bf P}}
\newcommand{\Q}{{\bf Q}}
\newcommand{\R}{\mathbb{R}}

\newcommand{\U}{{\bf U}}
\newcommand{\V}{{\bf V}}
\newcommand{\W}{{\bf W}}
\newcommand{\X}{{\bf X}}
\newcommand{\Y}{{\bf Y}}
\newcommand{\Z}{{\bf Z}}

\newcommand{\vareps}{\tau}

\renewcommand{\v}{{\bf v}}

\renewcommand{\P}{{\bf P}}

\newcommand{\Lam}{\boldsymbol{\Lambda}}
\newcommand{\Sig}{{\bf C}}

\newcommand{\ol}{\overline}

\newcommand{\bfP}{{\bf P}}

\title{A biologically plausible neural network for multi-channel Canonical Correlation Analysis}

\author[1]{David Lipshutz\thanks{Equal contribution}}
\author[1]{Yanis Bahroun\protect\footnotemark[1]}
\author[1]{Siavash Golkar\protect\footnotemark[1]}
\author[1,2]{\mbox{Anirvan M.\ Sengupta}}
\author[1,3]{Dmitri B.\ Chklovskii}

\affil[1]{Center for Computational Neuroscience, Flatiron Institute}
\affil[2]{Department of Physics and Astronomy, Rutgers University}
\affil[3]{Neuroscience Institute, NYU Medical Center}

\begin{document}

\maketitle

\begin{abstract}
Cortical pyramidal neurons receive inputs from multiple distinct neural populations and integrate these inputs in separate dendritic compartments.
We explore the possibility that cortical microcircuits implement Canonical Correlation Analysis (CCA), an unsupervised learning method that projects the inputs onto a common subspace so as to maximize the correlations between the projections. 
To this end, we seek a multi-channel CCA algorithm that can be implemented in a biologically plausible neural network.
For biological plausibility, we require that the network operates in the online setting and its synaptic update rules are local.
Starting from a novel CCA objective function, we derive an online optimization algorithm whose optimization steps can be implemented in a single-layer neural network with multi-compartmental neurons and local non-Hebbian learning rules.
We also derive an extension of our online CCA algorithm with adaptive output rank and output whitening.
Interestingly, the extension maps onto a neural network whose neural architecture and synaptic updates resemble neural circuitry and non-Hebbian plasticity observed in the cortex.
\end{abstract}

\clearpage

{
\hypersetup{linkcolor=black}
\tableofcontents
}

\clearpage

\section{Introduction}

Our brains can effortlessly extract a latent source contributing to two synchronous data streams, often from different sensory modalities. Consider, for example, following an actor while watching a movie with a soundtrack. We can easily pay attention to the actor’s gesticulation and voice while filtering out irrelevant visual and auditory signals. How can biological neurons accomplish such multi-sensory integration?

In this paper, we explore an algorithm for solving a linear version of this problem known as Canonical Correlation Analysis (CCA) \cite{hotelling1936relations}. In CCA, the two synchronous datasets, known as views, are projected onto a common lower-dimensional subspace so that the projections are maximally correlated. For simple generative models, the sum of these projections yields an optimal estimate of the latent source \cite{bach2005probabilistic}. CCA is a popular method because it has a closed form exact solution in terms of the Singular Value Decomposition (SVD) of the correlation matrix. Therefore, the projections can be computed using fast and well understood spectral numerical methods.

To serve as a viable model of a neuronal circuit, the CCA algorithm must map onto a neural network consistent with basic biological facts.  
For our purposes, we say that a network is ``biologically plausible'' if it satisfies the following two minimal requirements: (i) the network operates in the online setting, i.e., upon receiving an input, it computes the corresponding output without relying on the storage of any significant fraction of the full dataset, and (ii) the learning rules are local in the sense that each synaptic update depends only on the variables that are available as biophysical quantities represented in the pre- or post-synaptic neurons.

There are a number of neural network implementations of CCA \citep{lai1999neural,pezeshki2003network,gou2004canonical,via2007learning}; however, most of these networks use non-local learning rules and are therefore not biologically plausible.
One exception is the normative neural network model derived by \citet{Zhao2020}.
They start with formulating an objective for single-(output) channel CCA and derive an online optimization algorithm (previously proposed in \cite{lai1999neural}) that maps onto a pyramidal neuron with three electrotonic compartments: soma, as well as apical and basal dendrites.  The apical and basal synaptic inputs represent the two views, the two dendritic compartments extract highly correlated CCA projections of the inputs and the soma computes the sum of projections and outputs it downstream as action potentials. The communication between the compartments is implemented by calcium plateaus that also mediate non-Hebbian but local synaptic plasticity.

Whereas \citet{Zhao2020} also propose circuits of pyramidal neurons for multi-channel CCA their implementations lack biological plausibility. In one implementation, they resort to deflation where the circuit sequentially finds projections of the two views.  Implementing this algorithm in a neural network requires a centralized mechanism to facilitate the sequential updates, and there is no experimental evidence of such a biological mechanism. In another implementation that does not require a centralized mechanism, the neural network has asymmetric lateral connections among pyramidal neurons. However, that algorithm is not derived from a principled objective for CCA and the network architecture does not match the neuronal circuitry observed in cortical microcircuits.

\begin{figure}
    \centering
    \includegraphics[width=0.5\textwidth]{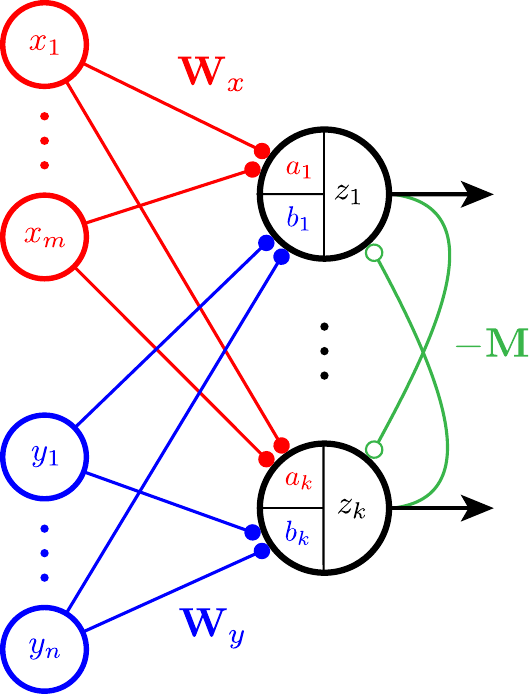}
    \caption{Single-layer network architecture with $k$ multi-compartmental neurons for outputting the sum of the canonical correlation subspace projections (CCSPs) $\z=(z_1,\dots,z_k)$, see Algorithm \ref{alg:online}. 
    Here $\a=\W_x\x$ and $\b=\W_y\y$ are projections of the views $\x=(x_1,\dots,x_m)$ and $\y=(y_1,\dots,y_n)$ onto a common $k$-dimensional subspace. 
    The output, $\z=\M^{-1}(\a+\b)$, is the sum of the CCSPs and is computed using recurrent lateral connections.
    The components of $\a$, $\b$ and $\z$ are represented in three separate compartments of the neurons.
    Filled circles denote non-Hebbian synapses and empty circles denote anti-Hebbian synapses.
    Importantly, each synaptic update depends only on variables represented locally.}
    \label{fig:ann}
\end{figure}

In this work, starting with a novel similarity-based CCA objective function, we derive a novel offline CCA algorithm (Algorithm \ref{alg:offline}) and an online multi-channel CCA algorithm (Algorithm \ref{alg:online}), which can be implemented in a single-layer network composed of three-compartment neurons and with local non-Hebbian synaptic update rules, Figure \ref{fig:ann}.  While our neural network implementation of CCA captures salient features of cortical microcircuits, the  network  includes  direct  lateral  connections between the principal neurons (see Figure \ref{fig:ann}), which is in contrast to cortical microcircuits where lateral influence between cortical pyramidal neurons is often indirect and mediated by interneurons.  With this in mind, we derive an extension of our on-line CCA algorithm (Algorithm \ref{alg:adaptive}), which adaptively chooses the rank of the output based on the level of correlation captured, and also whitens the output.  This extension is especially relevant for online unsupervised learning algorithms which are often confronted with the challenge of adapting to non-stationary input streams.  In addition, the algorithm naturally maps onto a neural network with multi-compartmental principal neurons and without direct lateral connections between the principal neurons (see Figure \ref{fig:1} of Section \ref{sec:bio}). 
Interestingly, both the neural architecture and local, non-Hebbian plasticity resemble neural circuitry and synaptic plasticity in cortical microcircuits.

There are a number of existing consequential models of cortical microcircuits with multi-compartmental neurons and non-Hebbian plasticity \citep{kording2001supervised,urbanczik2014learning,guerguiev2017towards,sacramento2018dendritic,haga2018dendritic,richards2019dendritic,milstein2020bidirectional}.
These models provide mechanistic descriptions of the neural dynamics and synaptic plasticity and account for many experimental observations, including the nonlinearity of neural outputs and the layered organization of the cortex.
While our neural network model is single-layered and linear, it is derived from a principled CCA objective function, which has several advantages.
First, since biological neural networks evolved to adaptively perform behaviorally relevant computations, it is natural to view them as optimizing a relevant objective function.
Second, our approach clarifies which features of the network (e.g., multi-compartmental neurons and non-Hebbian synaptic updates) are central to computing correlations.
Finally, since the optimization algorithm is derived from a CCA objective that can be solved offline, the neural activities and synaptic weights can be analytically predicted for any input without resorting to numerical simulation.
In this way, our neural network model is interpretable and analytically tractable, and provides a useful complement to nonlinear, layered neural network models.

\paragraph{Organization.}
The remainder of this work is organized as follows.
We state the CCA problem in Section~\ref{sec:cca}.
In Section \ref{sec:algorithms}, we introduce a novel objective for the CCA problem and derive offline and online CCA algorithms.
In Section~\ref{sec:extensions}, we derive an extension of our CCA algorithm, and in Section \ref{sec:bio}, we map the extension onto a simplified cortical microcircuit.
We provide results of numerical simulations in Section~\ref{sec:numerics}.

\paragraph{Notation.}
For positive integers $p,q$, let $\R^p$ denote $p$-dimensional Euclidean space, and let $\R^{p\times q}$ denote the set of $p\times q$ real-valued matrices equipped with the Frobenius norm $\|\cdot\|_\text{F}$.
We use boldface lower-case letters (e.g., $\v$) to denote vectors and boldface upper-case letters (e.g., $\M$) to denote matrices.
We let $O(p)$ denote the set of $p\times p$ orthogonal matrices and $\mathcal{S}_{++}^p$ denote the set of $p\times p$ positive definite matrices. 
We let $\I_p$ denote the $p\times p$ identity matrix.

\section{Canonical Correlation Analysis}
\label{sec:cca}

\begin{figure}
    \centering
    \includegraphics[width=\textwidth]{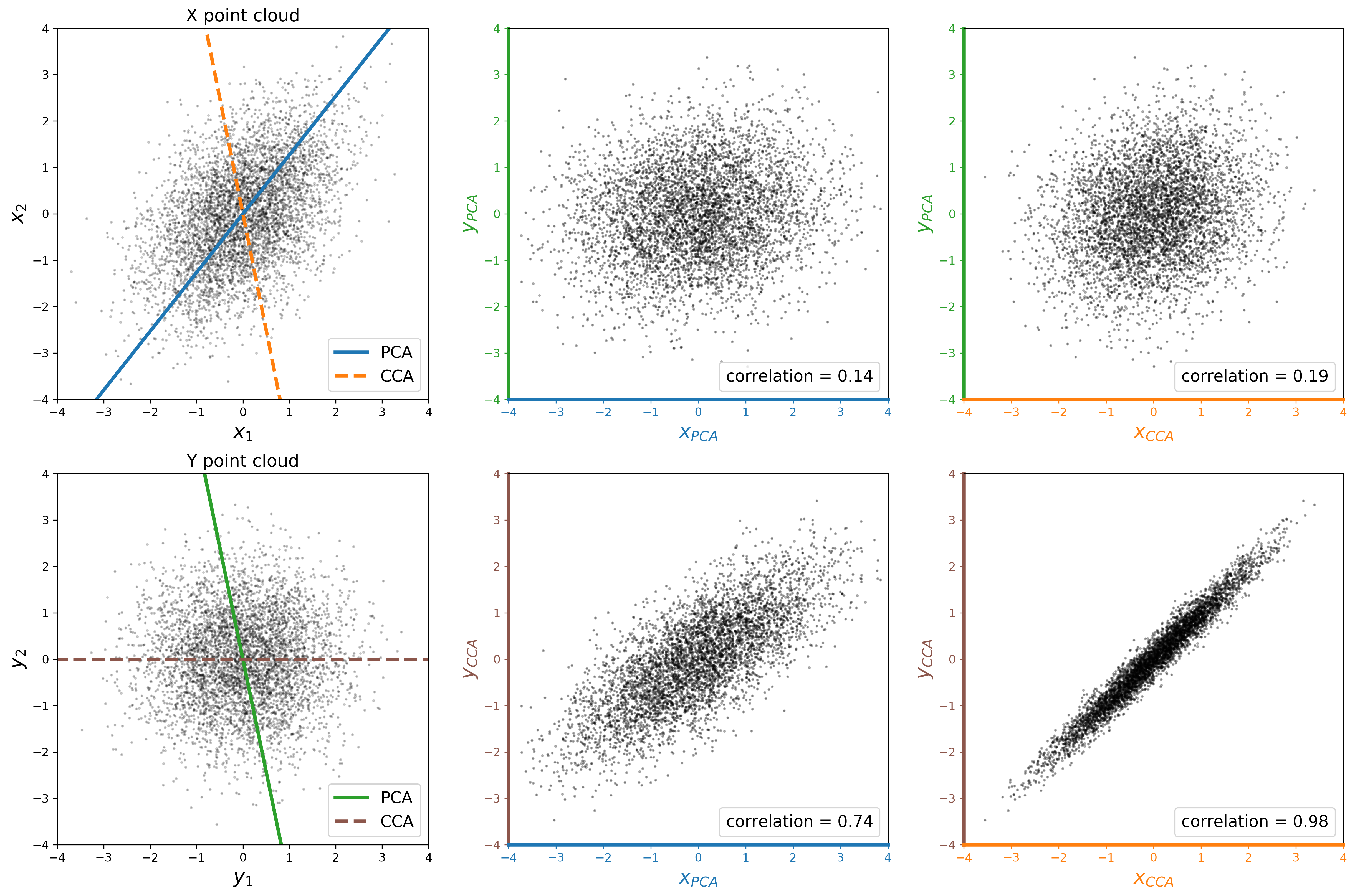}
    \caption{Illustration of CCA. The left column depicts point clouds of 2-dimensional views $\X$ and $\Y$ with lines indicating the span of the top principal component and top canonical correlation basis vector for each view (labeled ``PCA'' and ``CCA'', respectively). The center and right columns depict point clouds of joint 1-dimensional projections of $\X$ and $\Y$ onto their top principal component or top canonical correlation basis vector, with the correlations between the 2 projections listed. As illustrated in the lower right plot, the correlation between the projected views is maximized when each view is projected onto its top canonical correlation basis vector.}
    \label{fig:cca}
\end{figure}

Given $T$ pairs of full-rank, centered input data samples $(\x_1,\y_1),\dots,(\x_T,\y_T)\in\R^m\times\R^n$ and $k\le\min(m,n)$, the CCA problem is to find $k$-dimensional linear projections of the views $\x_1,\dots,\x_T$ and $\y_1,\dots,\y_T$ that are maximally correlated, see Figure \ref{fig:cca}.
To be precise, consider the CCA objective
\begin{align}
    \label{eq:cca1}
    \argmax{\V_x\in\R^{m\times k},\V_y\in\R^{n\times k}}\tr\left(\V_x^\top\Sig_{xy}\V_y\right)
\end{align}
subject to the whitening constraint\footnote{This constraint differs slightly from the usual CCA whitening constraint $\V_x^\top\Sig_{xx}\V_x=\V_y^\top\Sig_{yy}\V_y=\I_k$; however, the constraints are equivalent up to a scaling factor of 2.}
\begin{align}
    \label{eq:cca2}
    \V_x^\top\Sig_{xx}\V_x+\V_y^\top\Sig_{yy}\V_y=\I_k,
\end{align}
where we have defined the sample covariance matrices
\begin{align}\label{eq:cov1}
    \Sig_{xx}&:=\frac1T\sum_{t=1}^T\x_t\x_t^\top,&&\Sig_{xy}:=\frac1T\sum_{t=1}^T\x_t\y_t^\top,&&\Sig_{yy}:=\frac1T\sum_{t=1}^T\y_t\y_t^\top.
\end{align}

To compute the solution of the CCA objective \eqref{eq:cca1}--\eqref{eq:cca2}, define the $m\times n$ {\it correlation matrix}
\begin{align}
    \label{eq:corr}
    \C_{xy}:=\Sig_{xx}^{-1/2}\Sig_{xy}\Sig_{yy}^{-1/2}.
\end{align}
Let $\rho_1\ge\cdots\ge\rho_{\min(m,n)}$ denote the singular values, and let ${\bf U}_x\in O(m)$ and ${\bf U}_y\in O(n)$ denote the matrices whose column vectors are respectively the left- and right-singular vectors of the correlation matrix.
The $i$\textsuperscript{th} singular value $\rho_i$ is referred to as the $i$\textsuperscript{th} {\it canonical correlation}, and the $i$\textsuperscript{th} column vectors of $\Sig_{xx}^{-1/2}{\bf U}_x$ and $\Sig_{yy}^{-1/2}{\bf U}_y$ are jointly referred to as the $i$\textsuperscript{th} pair of {\it canonical correlation basis vectors}, for $i=1,\dots,\min(m,n)$.
The maximal value of the trace in Equation \eqref{eq:cca1} is the normalized sum of canonical correlations: $(\rho_1+\cdots+\rho_k)/2$.
For simplicity, we assume $\rho_k>\rho_{k+1}$ so the subspace spanned by the first $k$ canonical correlation basis vectors is unique.
In this case, every solution of the CCA objective \eqref{eq:cca1}--\eqref{eq:cca2}, denoted $(\ol\V_x,\ol\V_y)$, is of the form 
\begin{align}
\label{eq:Vbar}
    \ol\V_x=\Sig_{xx}^{-1/2}{\bf U}_x^{(k)}\Q,&&\ol\V_y=\Sig_{yy}^{-1/2}{\bf U}_y^{(k)}\Q,
\end{align}
where ${\bf U}_x^{(k)}$ (resp.\ ${\bf U}_y^{(k)}$) is the $m\times k$ (resp.\ $n\times k$) matrix whose $i$\textsuperscript{th} column vector is equal to the $i$\textsuperscript{th} column vector of ${\bf U}_x$ (resp.\ ${\bf U}_y$) for $i=1,\dots,k$, and $\Q\in O(k)$ is any orthogonal matrix.
Since the column vectors of any solution $(\ol\V_x,\ol\V_y)$ span the same subspaces as the first $k$ pairs of canonical correlation basis vectors, we refer to the column vectors of $\ol\V_x$ and $\ol\V_y$ as {\it basis vectors}.
We refer to the $k$-dimensional projections $\ol\V_x^\top\x_t$ and $\ol\V_y^\top\y_t$ as {\it canonical correlation subspace projections (CCSPs)}.

The focus of this work is to derive a single-layer biologically plausible network whose input at each time $t$ is the pair $(\x_t,\y_t)$ and the output is the following sum of the CCSPs: 
\begin{align}\label{eq:zt}
    \z_t:=\ol\V_x^\top\x_t+\ol\V_y^\top\y_t,
\end{align}
which, as mentioned in the introduction, is a highly relevant statistic (see, also, Section \ref{sec:datasets}).
This is in contrast to many existing CCA networks which output one or both CCSPs \citep{lai1999neural,pezeshki2003network,gou2004canonical,via2007learning,golkar2020}.
The components of the two input vectors $\x_t$ and $\y_t$ are represented by the activity of upstream neurons belonging to two different populations, which are integrated in separate compartments of the principal neurons in our network. 
The components of the output vector $\z_t$ are represented by the activity of the principal neurons in our network, see Figure~\ref{fig:ann}.

While CCA is typically viewed as an \textit{unsupervised} learning method, it can also be interpreted as a special case of the \textit{supervised} learning method Reduced-Rank Regression, in which case one input is the feature vector and the other input is the label (see, e.g., page 38 of \citep{velu2013multivariate}).
With this supervised learning view of CCA, the natural output of a CCA network is the CCSP of the feature vector.
In separate work \citep{golkar2020}, we derive an algorithm for the general Reduced-Rank Regression problem, which includes CCA as a special case, for outputting the projection of the feature vector.
The algorithm derived in \citep{golkar2020} resembles the adaptive CCA with output whitening algorithm that we derive in Section \ref{sec:extensions} of this work (see Algorithm \ref{alg:adaptive} as well as Appendix \ref{apdx:compare} for a detailed comparison of the two algorithms); however, there are significant advantages to the algorithm derived here. First, our network outputs the (whitened) sum of the CCSPs, which, as discussed above, is a relevant statistic in applications. The algorithm in \citep{golkar2020} only outputs the CCSP of the feature vector, which is natural when viewing CCA as a supervised learning method, but not when viewing CCA as an unsupervised learning method for integrating multi-view inputs. Second, in contrast to the algorithm derived in \citep{golkar2020}, our adaptive CCA with output whitening algorithm allows for adaptive output rank.
This is particularly important for analyzing non-stationary input streams, a challenge that brains regularly face.

\section{A biologically plausible CCA algorithm}
\label{sec:algorithms}

To derive a network that computes the sums of CCSPs for arbitrary input datasets, we adopt a normative approach in which we identify an appropriate cost function whose optimization leads to an online algorithm that can be implemented by a network with local learning rules.
Previously, such approach was taken to derive a biologically plausible PCA network from a similarity matching objective function \citep{pehlevan2015hebbian}. 
We leverage this work by reformulating a CCA problem in terms of PCA of a modified dataset and then solving it using similarity matching.

\subsection{A similarity matching objective}\label{sec:sm}

First, we note that the sums of CCSPs $\z_1,\dots,\z_T$ are equal to the principal subspace projections of the data $\boldsymbol{\xi}_1,\dots,\boldsymbol{\xi}_T$, where $\boldsymbol{\xi}_t$ is the following $d$-dimensional vector of concatenated whitened inputs (recall $d:=m+n$):
\begin{align}
\label{eq:xi}
    \boldsymbol{\xi}_t:=
    \begin{bmatrix}
        \Sig_{xx}^{-1/2}\x_t \\
        \Sig_{yy}^{-1/2}\y_t
    \end{bmatrix}.
\end{align}
(See Appendix \ref{apdx:sm} for a detailed justification.)
Next, we use the fact that the principal subspace projections can be expressed in terms of solutions of similarity matching objectives. 
To this end, we define the matrices $\boldsymbol{\Xi}:=[\boldsymbol{\xi}_1,\dots,\boldsymbol{\xi}_T]\in\R^{d\times T}$ and $\Z:=[\z_1,\dots,\z_T]\in\R^{k\times T}$, so that $\Z$ is a linear projection of $\boldsymbol{\Xi}$ onto its $k$-dimensional principal subspace.
As shown in \citep{cox2000multidimensional,williams2001connection}, the principal subspace projection $\Z$ is a solution of following similarity matching objective:
\begin{align}
\label{eq:sm}
    \argmin{\Z\in\R^{k\times T}}\frac{1}{2T^2}\norm{\Z^\top\Z-\boldsymbol{\Xi}^\top\boldsymbol{\Xi}}_\text{F}^2
    .
\end{align}
The objective \eqref{eq:sm}, which comes from classical multidimensional scaling \citep{cox2000multidimensional}, minimizes the difference between the similarity of output pairs, $\z_t^\top\z_{t'}$, and the similarity of input pairs, $\boldsymbol{\xi}_t^\top\boldsymbol{\xi}_{t'}$, where similarity is measured in terms of inner products.
Finally, defining $\X:=[\x_1,\dots,\x_T]\in\R^{m\times T}$ and $\Y:=[\y_1,\dots,\y_T]\in\R^{n\times T}$, we use the definition of $\boldsymbol{\xi}_t$ in Equation \eqref{eq:xi} to rewrite the objective \eqref{eq:sm} as follows:
\begin{align}
\label{eq:similarity}
    \argmin{\Z\in\R^{k\times T}}\frac{1}{2T^2}\norm{\Z^\top\Z-\X^\top\Sig_{xx}^{-1}\X-\Y^\top\Sig_{yy}^{-1}\Y}_\text{F}^2.
\end{align}
In the remainder of this section, we derive our online CCA algorithm.
Then, in Sections \ref{sec:extensions} and \ref{sec:bio}, we derive an extension of our CCA algorithm and map it onto the cortical microcircuit.
The reader who is primarily interested in the derivation of the extension and its relation to the cortical microcircuit can safely skip to Section \ref{sec:extensions}.

\subsection{A min-max objective}
\label{sec:minmax}

While the similarity matching objective \eqref{eq:similarity} can be minimized by taking gradient descent steps with respect to $\Z$, this would not lead to an online algorithm because such computation requires combining data from different time steps.
Rather, we introduce auxiliary matrix variables, which store sufficient statistics allowing for the CCA computation using solely contemporary inputs and will correspond to synaptic weights in the network implementation, and rewrite the minimization problem \eqref{eq:similarity} as a min-max problem.

Expanding the square in Equation~\eqref{eq:similarity} and dropping terms that do not depend on $\Z$ yields the minimization problem
\begin{align*}
    &\min_{\Z\in\R^{k\times T}}-\frac{1}{T^2}\tr\left(\Z^\top\Z\X^\top\Sig_{xx}^{-1}\X\right)-\frac{1}{T^2}\tr\left(\Z^\top\Z\Y^\top\Sig_{yy}^{-1}\Y\right)+\frac{1}{2T^2}\tr\left(\Z^\top\Z\Z^\top\Z\right).
\end{align*}
Next, we introduce dynamic matrix variables $\W_x$, $\W_y$ and $\M$ in place of the matrices $\frac{1}{T}\Z\X^\top\Sig_{xx}^{-1}$, $\frac{1}{T}\Z\Y^\top\Sig_{yy}^{-1}$ and $\frac1T\Z\Z^\top$, respectively, and rewrite the minimization problem as a min-max problem:
\begin{align*}
    &\min_{\Z\in\R^{k\times T}}\min_{\W_x\in\R^{k\times m}}\min_{\W_y\in\R^{k\times n}}\max_{\M\in\mathcal{S}_{++}^k}L(\W_x,\W_y,\M,\Z)\\
\end{align*}
where
\begin{multline}
    \label{eq:L}
    L(\W_x,\W_y,\M,\Z):=\frac1T\tr\left(-2\Z^\top\W_x\X-2\Z^\top\W_y\Y+\Z^\top\M\Z\right)\\
    +\tr\left(\W_x\Sig_{xx}\W_x^\top+\W_y\Sig_{yy}\W_y^\top-\frac12\M^2\right).
\end{multline}
To verify the above substitutions are valid, it suffices to optimize over the matrices $\W_x$, $\W_y$ and $\M$; e.g., by differentiating $L(\W_x,\W_y,\M,\Z)$ with respect to $\W_x$, $\W_y$ or $\M$, setting the derivative equal to zero, and solving for $\W_x$, $\W_y$ or $\M$.
Finally, we interchange the order of minimization with respect to $\Z$ and $(\W_x,\W_y)$, as well as the order of minimization with respect to $\Z$ and maximization with respect to $\M$:
\begin{align}\label{eq:minmax}
    &\min_{\W_x\in\R^{k\times m}}\min_{\W_y\in\R^{k\times n}}\max_{\M\in\mathcal{S}_{++}^k}\min_{\Z\in\R^{k\times T}}L(\W_x,\W_y,\M,\Z).
\end{align}
The second interchange is justified by the fact that $L(\W_x,\W_y,\M,\Z)$ satisfies the saddle point property with respect to $\Z$ and $\M$, which follows from its strict convexity in $\Z$ (since $\M$ is positive definite) and strict concavity in $\M$.

Given an optimal quadruple of the min-max problem \eqref{eq:minmax}, we can compute the basis vectors, as follows.
First, minimizing the objective $L(\W_x,\W_y,\M,\Z)$ over $\Z$ yields the relation
\begin{align}
\label{eq:Zoptimal}
    \ol\Z:=\argmin{\Z\in\R^{k\times T}}L(\W_x,\W_y,\M,\Z)=\M^{-1}\W_x\X+\M^{-1}\W_y\Y.
\end{align}
Therefore, if $(\ol\W_x,\ol\W_y,\ol\M,\ol\Z)$ is an optimal quadruple of the min-max problem \eqref{eq:minmax}, it follows from Equation~\eqref{eq:zt} that the corresponding basis vectors satisfy
\begin{align}\label{eq:VxVy}
    \ol\V_x^\top=\ol\M^{-1}\ol\W_x\qquad\text{and}\qquad\ol\V_y^\top=\ol\M^{-1}\ol\W_y.
\end{align}

\subsection{An offline CCA algorithm}

Before deriving our online CCA algorithm, we first demonstrate how the objective \eqref{eq:minmax} can be optimized in the offline setting, where one has access to the data matrices $\X$ and $\Y$ in their entirety.
In this case, the algorithm solves the min-max problem \eqref{eq:minmax} by alternating minimization and maximization steps.
First, for fixed $\W_x$, $\W_y$ and $\M$, we minimize the objective function $L(\W_x,\W_y,\M,\Z)$ over $\Z$ to obtain the minimum $\ol\Z$ defined in Equation~\eqref{eq:Zoptimal}.
Then, with $\ol\Z$ fixed, we perform a gradient descent-ascent step with respect to $(\W_x,\W_y)$ and $\M$:
\begin{align*}
    \W_x&\gets\W_x-\eta\frac{\partial L(\W_x,\W_y,\M,\ol\Z)}{\partial \W_x}\\
    \W_x&\gets\W_y-\eta\frac{\partial L(\W_x,\W_y,\M,\ol\Z)}{\partial \W_y}\\
    \M&\gets\M+\frac{\eta}{\tau}\frac{\partial L(\W_x,\W_y,\M,\ol\Z)}{\partial \M}.
\end{align*}
Here $\eta>0$ is the learning rate for $\W_x$ and $\W_y$, which may depend on the iteration, and $\vareps>0$ is the ratio of the learning rates for $\W_x$ (or $\W_y$) and $\M$.
Substituting in the explicit expressions for the partial derivatives of $L(\W_x,\W_y,\M,\ol\Z)$ yields our offline CCA algorithm (Algorithm~\ref{alg:offline}), which we refer to as Offline-CCA.

\begin{algorithm}
  \caption{Offline-CCA}
  \label{alg:offline}
\begin{algorithmic}
  \STATE {\bfseries input:} data matrices $\X$, $\Y$; dimension $k$
  \STATE {\bfseries initialize:} matrices $\W_x$, $\W_y$ and positive definite matrix $\M$ 
  \STATE $\Sig_{xx}\gets{\frac1T}\X\X^\top$ \quad ; \quad  $\Sig_{yy}\gets{\frac1T}\Y\Y^\top$ \hspace{91pt} $\triangleright\;$ covariance matrices 
  \REPEAT
  \STATE $\Z\gets\M^{-1}\W_x\X+\M^{-1}\W_x\Y$ \hspace{106pt} $\triangleright\;$ optimize over output
  \STATE $\W_x \gets \W_x + 2\eta\left(\frac1T\Z\X^\top-\W_x\Sig_{xx}\right)$ \hspace{30pt} $\triangleright\;$ gradient descent-ascent steps
  \STATE $\W_y \gets \W_y + 2\eta\left(\frac1T\Z\Y^\top-\W_y\Sig_{yy}\right)$
  \STATE $\M \gets \M + \frac\eta\vareps\left(\frac1T\Z\Z^\top-\M\right)$ 
  \UNTIL{convergence}
\end{algorithmic}
\end{algorithm}

Recall that $\M$ is optimized over the set of positive definite matrices $\mathcal{S}_{++}^k$. To ensure that $\M$ remains positive definite after each update, note that the update rule for $\M$ can be rewritten as the following convex combination (provided $\eta\le\tau$): $\M\gets(1-\frac\eta\tau)\M+\frac\eta\tau(\frac1T\Z\Z^\top)$. Since $\frac1T\Z\Z^\top$ is positive semidefinite, to guarantee that $\M$ remains positive definite given a positive definite initialization, it suffices to assume that $\eta<\vareps$.

\subsection{An online CCA algorithm}
\label{sec:online}

In the online setting, the input data $(\x_t,\y_t)$ are streamed one at a time and the algorithm must compute its output $\z_t$ without accessing any significant fraction of $\X$ and $\Y$. 
To derive an online algorithm, it is useful to write the cost function as an average over time-separable terms:
\begin{align*}
    L(\W_x,\W_y,\M,\Z)=\frac1T\sum_{t=1}^Tl_t(\W_x,\W_y,\M,\z_t),
\end{align*}
where
\begin{multline}
\label{eq:lt}
    l_t(\W_x,\W_y,\M,\z_t):=-2\z_t^\top\W_x\x_t-2\z_t^\top\W_y\y_t+\z_t^\top\M\z_t\\
    +\tr\left(\W_x\x_t\x_t^\top\W_x^\top+\W_y\y_t\y_t^\top\W_y^\top-\frac12\M^2\right).
\end{multline}
At iteration $t$, to compute the output $\z_t$, we minimize the cost function $l_t(\W_x,\W_y,\M,\z_t)$ with respect to $\z_t$ by running the following gradient descent dynamics to equilibrium:
\begin{align}
\label{eq:dzdgamma}
    \frac{d\z_t(\gamma)}{d\gamma}=\a_t+\b_t-\M\z_t(\gamma),
\end{align}
where we have defined the following $k$-dimensional projections of the inputs: $\a_t:=\W_x\x_t$ and $\b_t:=\W_y\y_t$.
These dynamics, which will correspond to recurrent neural dynamics in our network implementation, are assumed to occur on a fast timescale, allowing $\z_t(\gamma)$ to equilibrate at $\ol\z_t:=\M^{-1}(\a_t+\b_t)$ before the algorithm outputs its value.
After $\z_t(\gamma)$ equilibrates, we update the matrices $(\W_x,\W_y,\M)$ by taking a stochastic gradient descent-ascent step of the cost function $l_t(\W_x,\W_y,\M,\ol\z_t)$ with respect to $(\W_x,\W_y)$ and $\M$:
\begin{align*}
    \W_x&\gets\W_x-\eta\frac{\partial l_t(\W_x,\W_y,\M,\ol\z_t)}{\partial \W_x}\\
    \W_x&\gets\W_y-\eta\frac{\partial l_t(\W_x,\W_y,\M,\ol\z_t)}{\partial \W_y}\\
    \M&\gets\M+\frac{\eta}{\tau}\frac{\partial l_t(\W_x,\W_y,\M,\ol\z_t)}{\partial \M}.
\end{align*}
Substituting in the explicit expressions for the partial derivatives of $l_t(\W_x,\W_y,\M,\ol\z_t)$ yields our online CCA algorithm (Algorithm~\ref{alg:online}), which we refer to as Bio-CCA.

\begin{algorithm}[ht]
  \caption{Bio-CCA}
  \label{alg:online}
\begin{algorithmic}
  \STATE {\bfseries input} data $\{(\x_1,\y_1),\dots,(\x_T,\y_T)\}$; dimension $k$
  \STATE {\bfseries initialize} matrices $\W_x$, $\W_y$, and positive definite matrix $\M$.
  \FOR{$t=1, 2,\dots,T $}
  \STATE $\a_t\gets\W_x\x_t\quad;\quad\b_t\gets\W_y\y_t$ \hspace{112pt} $\triangleright\;$ projection of inputs
  \STATE {\bfseries run}
  \STATE \Indp $\frac{d\z_t(\gamma)}{d\gamma}=\a_t+\b_t-\M\z_t(\gamma)$ \hspace{127pt} $\triangleright\;$ neural dynamics
  \STATE \Indm {\bfseries until convergence}
  \STATE $\W_x \gets \W_x + 2\eta ( \z_t - \a_t )\x_t^\top$ \hspace{131pt} $\triangleright\;$ synaptic updates
  \STATE $\W_y \gets \W_y + 2\eta ( \z_t - \b_t )\y_t^\top$
  \STATE $\M \gets \M + \frac{\eta}{\vareps} (\z_t \z_t^\top-\M)$
  \ENDFOR
\end{algorithmic}
\end{algorithm}

Algorithm~\ref{alg:online} can be implemented in a biologically plausible single-layer network with $k$ neurons that each consist of three separate compartments, Figure~\ref{fig:ann}.
At each time step, the inputs $\x_t$ and $\y_t$ are multiplied by the respective feedforward synapses $\W_x$ and $\W_y$ to yield the $k$-dimensional vectors $\a_t$ and $\b_t$, which are represented in the first two compartments of the $k$ neurons.
Lateral synapses, $-\M$, connect the $k$ neurons.
The vector of neuronal outputs, $\z_t$, equals the normalized sum of the CCSPs, and is computed locally using recurrent dynamics in Equation~\eqref{eq:dzdgamma}.
The synaptic updates can be written elementwise, as follows:
\begin{align*}
    W_{x,ij}&\gets W_{x,ij}+\eta(z_{t,i}-a_{t,i})x_{t,j},&&1\le i\le k,\;1\le j\le m,\\
    W_{y,ij}&\gets W_{y,ij}+\eta(z_{t,i}-b_{t,i})y_{t,j},&&1\le i\le k,\;1\le j\le n,\\
    M_{ij}&\gets M_{ij}+\frac\eta\vareps(z_{t,i}z_{t,j}-M_{ij}),&&1\le i,j\le k.
\end{align*}
As shown above, the update to synapse $W_{x,ij}$ (resp.\ $W_{y,ij}$), which connects the $j$\textsuperscript{th} input $x_{t,j}$ (resp.\ $y_{t,j}$) to the $i$\textsuperscript{th} output neuron, depends only on the quantities $z_{t,i}$, $a_{t,i}$ (resp.\ $b_{t,i}$), and $x_{t,j}$ (resp.\ $y_{t,j}$), which are represented in the pre- and post-synaptic neurons, so the updates are local, but non-Hebbian due to the contribution from the $a_{t,i}$ (resp.\ $b_{t,i}$) term.
Similarly, the update to synapse $-M_{ij}$, which connects the $j$\textsuperscript{th} output neuron to the $i$\textsuperscript{th} output neuron, is inversely proportional to $z_{t,i}z_{t,j}$, the product of the outputs of the pre- and post-synaptic neurons, so the updates are local and anti-Hebbian.

\section{Online adaptive CCA with output whitening}\label{sec:extensions}

We now introduce an extension of Bio-CCA which addresses two biologically relevant issues.
First, Bio-CCA a priori sets the output rank at $k$; however, it may be advantageous for a neural circuit to instead adaptively set the output rank depending on the level of correlation captured.
In particular, this can be achieved by projecting each view onto the subspace spanned by the canonical correlation basis vectors which correspond to canonical correlations that exceed a threshold.
Second, it is useful from an information theoretic perspective for neural circuits to whiten their outputs \citep{plumbley1993hebbian}, and there is experimental evidence that neural outputs in the cortex are decorrelated \citep{ecker2010decorrelated,miura2012odor}.
Both adaptive output rank and output whitening modifications were implemented for a PCA network by \citet{pehlevan2015normative}, and can be adapted to the CCA setting.
Here we present the modifications without providing detailed proofs, which can be found in the supplement of \citep{pehlevan2015normative}.

In order to implement these extensions, we need to appropriately modify the similarity matching objective function~\eqref{eq:similarity}.
First, to adaptively choose the output rank, we add a quadratic penalty $\tr(\Z^\top\Z)$ to the objective function~\eqref{eq:similarity}:
\begin{equation}
    \label{eq:power_constraint}
    \argmin{\Z\in\R^{k\times T}}\frac{1}{2T^2}\norm{\Z^\top\Z-\X^\top\Sig_{xx}^{-1}\X-\Y^\top\Sig_{yy}^{-1}\Y}_\text{F}^2+\frac{\alpha}{T}\tr\left(\Z^\top\Z\right).
\end{equation}
The effect of the quadratic penalty is to rank constrain the output, with $\alpha\ge0$ acting as a threshold parameter on the eigenvalues values of the output covariance.

Next, to whiten the output, we expand the square in Equation~\eqref{eq:power_constraint} and replace the quartic term $\tr(\Z^\top\Z\Z^\top\Z)$ by a Lagrange constraint enforcing $\Z^\top\Z \preceq T \I_T$ (i.e., $T\I_T-\Z^\top\Z$ is positive semi-definite):
\begin{align}
\label{eq:whitening}
    \argmin{\Z\in\R^{k\times T}}\max_{\N\in\R^{k\times T}} &\frac{1}{T^2}\tr(-\Z^\top\Z\X^\top\Sig_{xx}^{-1}\X-\Z^\top\Z\Y^\top\Sig_{yy}^{-1}\Y+\alpha T\Z^\top\Z)\\ \notag
    &+\frac{1}{T^2}\tr[\N^\top \N(\Z^\top\Z - T\I_T)].
\end{align}
The effect of the Lagrange constraint in Equation~\eqref{eq:whitening} is to enforce that all non-zero eigenvalues of the output covariance are set to one.

Solutions of the objective \eqref{eq:whitening} can be expressed in terms of the eigendecomposition of the Gram matrix $\boldsymbol{\Xi}^\top\boldsymbol{\Xi}=T\U_\xi\Lam_\xi\U_\xi^\top$, where $\U_\xi\in O(T)$ is a matrix of eigenvectors and $\Lam_\xi=\text{diag}(\lambda_1,\dots,\lambda_d,0,\dots,0)$ is the $T\times T$ diagonal matrix whose non-zero entries $\lambda_1\ge\cdots\ge\lambda_d>0$ are the eigenvalues of the $d\times d$ covariance matrix 
\begin{align}
\label{eq:Sigxi}
    \Sig_{\xi\xi}:=\frac1T\sum_{t=1}^T\boldsymbol{\xi}_t\boldsymbol{\xi}_t^\top=\begin{bmatrix}
        \I_m & \C_{xy}\\
        \C_{xy}^\top & \I_n
    \end{bmatrix}.
\end{align}
Assume, for technical purposes, that $\alpha\not\in\{\lambda_1,\dots,\lambda_d\}$.
Then, as shown in \cite[Theorem 3]{pehlevan2015hebbian}, every solution $\widehat\Z$ of objective \eqref{eq:whitening} is of the form
\begin{align*}
    \widehat\Z=\Q\sqrt{T\widehat\Lam_\xi^{(k)}}{\U_\xi^{(k)}}^\top,&&\widehat\Lam_\xi^{(k)}=\text{diag}(H(\lambda_1-\alpha),\dots,H(\lambda_k-\alpha)),
\end{align*}
where $\Q\in O(k)$ is any orthogonal matrix, $\U_\xi^{(k)}\in\R^{T\times k}$ is the $T\times k$ matrix whose $i$\textsuperscript{th} column vector is equal to the $i$\textsuperscript{th} column vector of $\U_\xi$, for $i=1,\dots,k$, and $H$ is the Heaviside step function defined by $H(r)=1$ if $r>0$ and $H(r)=0$ otherwise.
Finally, we note that, in view of Equation \eqref{eq:Sigxi} and the SVD of $\C_{xy}$, the top $\min(m,n)$ eigenvalues of $\Sig_{\xi\xi}$ satisfy 
\begin{align*}
    \lambda_i=1+\rho_i,\qquad i=1,\dots,\min(m,n),
\end{align*}
where we recall that $\rho_1,\dots,\rho_{\min(m,n)}$ are the canonical correlations.
Thus, $H(\lambda_i-\alpha)=H(\rho_i-(\alpha-1))$, for $i=1,\dots,k$.
In other words, the objective~\eqref{eq:whitening} outputs the sum of the projections of the inputs $\x_t$ and $\y_t$ onto the canonical correlation subspace spanned by the (at most $k$) pairs of canonical correlation basis vectors associated with canonical correlations exceeding the threshold $\max(\alpha-1,0)$, and sets the non-zero output covariance eigenvalues to one, thus implementing both the adaptive output rank and output whitening modifications.

With the modified objective \eqref{eq:whitening} in hand, the next step is to derive an online algorithm.
Similar to Section~\ref{sec:minmax}, we introduce dynamic matrix variables $\W_x$, $\W_y$ and $\bfP$ in place of $\frac1T\Z\X^\top\Sig_{xx}^{-1}$, $\frac1T\Z\Y^\top\Sig_{yy}^{-1}$ and $\frac1T\Z\N^\top$ to rewrite the objective~\eqref{eq:whitening} as follows:
\begin{align}
\label{eq:minmaxtildeL}
    \argmin{\Z\in\R^{k\times T}}\max_{\N\in\R^{k\times T}}\min_{\W_x\in\R^{k\times m}}\min_{\W_y\in\R^{k\times n}}\max_{\bfP\in\R^{k\times k}} \widetilde{L}(\W_x,\W_y,\bfP,\Z,\N),
\end{align}
where
\begin{align*}
    \widetilde{L}(\W_x,\W_y,\bfP,\Z,\N)&:=\frac{1}{T}\tr\left(-2\Z^\top\W_x\X-2\Z^\top\W_y\Y+\alpha \Z^\top\Z\right)\\
    &\qquad+\frac1T\tr\left(2\N^\top \bfP^\top\Z - \N^\top\N\right)\\
    &\qquad+\tr\left(\W_x\Sig_{xx}\W_x^\top+\W_y\Sig_{yy}\W_y^\top-\bfP\bfP^\top\right).
\end{align*}
After interchanging the order of optimization, we solve the min-max optimization problem by taking online gradient descent-ascent steps, with descent step size $\eta$ and ascent step size $\frac\eta\vareps$.
Since the remaining steps are similar to those taken in Section~\ref{sec:algorithms} to derive Bio-CCA, we defer the details to Appendix \ref{apdx:extension} and simply state the online algorithm (Algorithm~\ref{alg:adaptive}), which we refer to as Adaptive Bio-CCA with output whitening.

\begin{algorithm}[ht]
  \caption{Adaptive Bio-CCA with output whitening}
  \label{alg:adaptive}
\begin{algorithmic}
  \STATE {\bfseries input} data $\{(\x_1,\y_1),\dots,(\x_T,\y_T)\}$; max output-dimension $k$; threshold $\alpha$
  \STATE {\bfseries initialize} weight matrices $\W_x$, $\W_y$, and $\bfP$.
  \FOR{$t=1, 2,\dots,T $}
  \STATE $\a_t\gets\W_x\x_t\quad;\quad\b_t\gets\W_y\y_t$ \hspace{111pt} $\triangleright\;$ projection of inputs
  \STATE {\bfseries run} 
  \STATE \Indp $\frac{d\z_t(\gamma)}{d\gamma}=\a_t+\b_t-\bfP\n_t(\gamma)-\alpha\z_t(\gamma)$ \hspace{81pt} $\triangleright\;$ neural dynamics 
  \STATE $\frac{d\n_t(\gamma)}{d\gamma}=\bfP^\top\z_t(\gamma)-\n_t(\gamma)$
  \STATE \Indm {\bfseries until convergence} 
  \STATE $\W_x \gets \W_x + \eta ( \z_t - \a_t )\x_t^\top$ \hspace{140pt}$\triangleright\;$ synaptic updates 
  \STATE $\W_y \gets \W_y + \eta ( \z_t - \b_t )\y_t^\top$ 
  \STATE $\bfP \gets \bfP + \frac{\eta}{\vareps} (\z_t \n_t^\top-\bfP) $
  \ENDFOR
\end{algorithmic}
\end{algorithm}

\section{Relation to cortical microcircuits}
\label{sec:bio}

We now show that Adaptive Bio-CCA with output whitening (Algorithm~\ref{alg:adaptive}) maps onto a neural network with local, non-Hebbian synaptic update rules that emulate salient aspects of synaptic plasticity found experimentally in cortical microcircuits (both in the neocortex and the hippocampus).

\begin{figure}
    \centering
    \includegraphics[width=.9\textwidth]{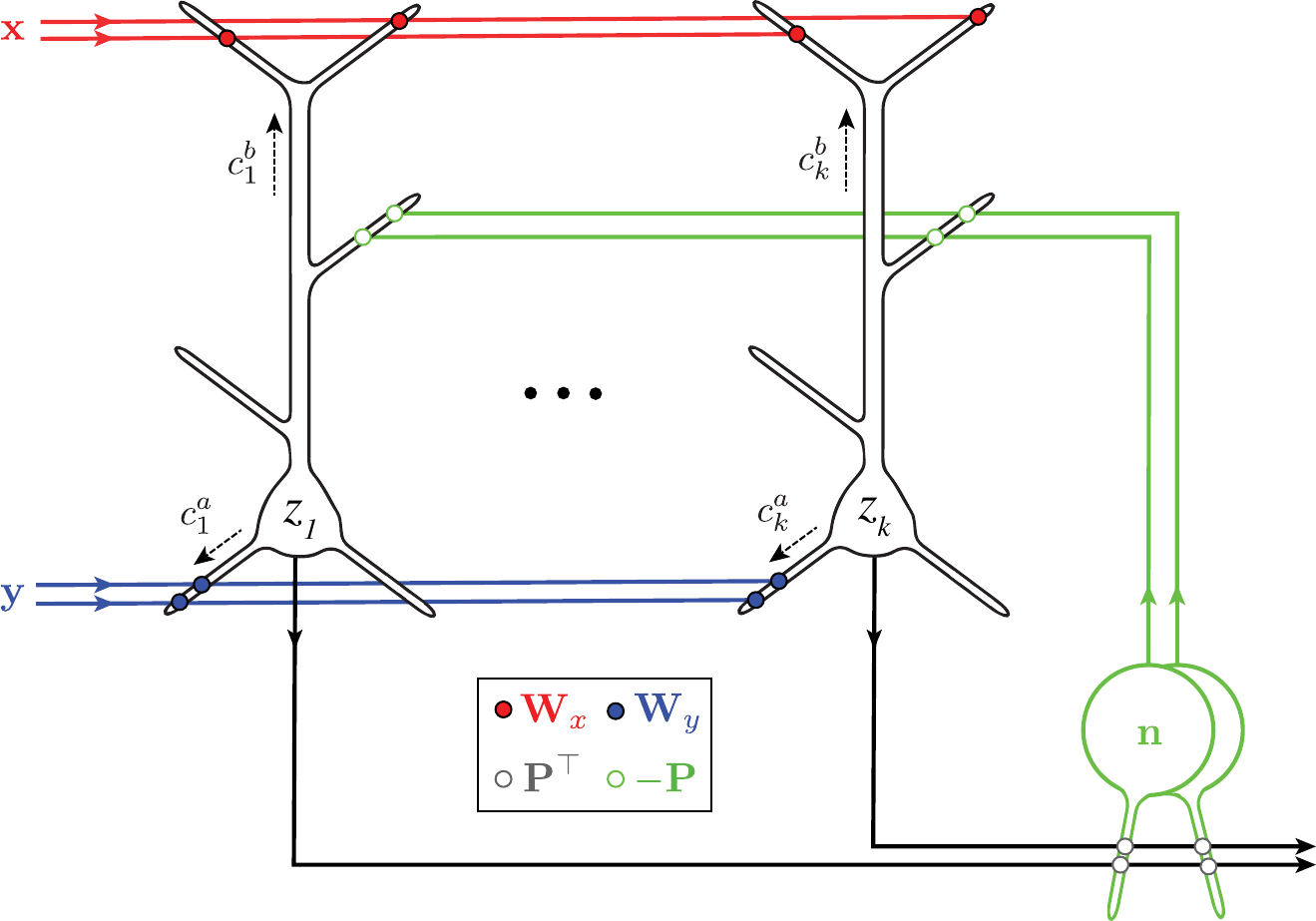}
    \vspace{20pt}
    \caption{Cortical microcircuit implementation of Adaptive Bio-CCA with output whitening (Algorithm~\ref{alg:adaptive}). 
    The black cell bodies denote pyramidal neurons, with the apical tufts pointing upwards. 
    The red and blue lines denote the axons respectively transmitting the apical input $\x$ and basal input $\y$.
    The black lines originating from the bases of the pyramidal neurons are their axons, which transmit their output $\z$. 
    The green cell bodies denote the interneurons and the green lines are their axons, which transmit their output ${\bf n}$.
    Filled circles denote non-Hebbian synapses whose updates are proportional to the input (i.e., $\x$ or $\y$) and the weighted sum of the calcium plateau potential plus backpropagating somatic output [i.e., $\c^b+(1-\alpha)\z$ or $\c^a+(1-\alpha)\z$].
    The directions of travel of these weighted sums are depicted using dashed lines with arrows.
    Empty circles denote Hebbian or anti-Hebbian synapses whose updates are proportional or inversely proportional to the pre- and post-synaptic activities.}
    \label{fig:1}
\end{figure}

Cortical microcircuits contain two classes of neurons: excitatory pyramidal neurons and inhibitory interneurons. 
Pyramidal neurons receive excitatory synaptic inputs from two distinct sources via their apical and basal dendrites.
The apical dendrites are all oriented in a single direction and the basal dendrites branch from the cell body in the opposite direction \citep{takahashimagee,Larkum2013}, Figure~\ref{fig:1}. 
The excitatory synaptic currents in the apical and basal dendrites are first integrated separately in their respective compartments \citep{takahashimagee,Larkum2013}. 
If the integrated excitatory current in the apical compartment exceeds the corresponding inhibitory input (the source of which is explained below) it produces a calcium plateau potential that propagates through the basal dendrites, driving plasticity \citep{takahashimagee,Larkum2013,bittner2015}. 
When the apical calcium plateau potential and basal dendritic current coincidentally arrive in the soma, they generate a burst in spiking output \citep{larkum1999new,Larkum2013, bittner2015}.
Inhibitory interneurons integrate pyramidal outputs and reciprocally inhibit the apical dendrites of pyramidal neurons, thus closing the loop. 

We propose that a network of $k$ pyramidal neurons implements CCA on the inputs received by apical and basal dendrites and outputs the whitened sum of CCSPs (Algorithm~\ref{alg:adaptive}). 
In our model, each pyramidal neuron has three compartments --- two compartments for the apical and basal dendritic currents, and one compartment for the somatic output.
The two datasets $\X$ and $\Y$ are represented as activity vectors $\x_t$ and $\y_t$ streamed onto the apical and basal dendrites respectively, Figure~\ref{fig:1}. 
At each time step, the activity vectors are multiplied by the corresponding synaptic weights to yield localized apical and basal dendritic currents, $\a_t=\W_x\x_t$ and $\b_t=\W_y\y_t$, thus implementing projection onto the common subspace. 
This is followed by the following linear recurrent neural dynamics:
\begin{align}\label{eq:dzt}
    \frac{d\z_t(\gamma)}{d\gamma}&=\a_t+\b_t-\bfP\n_t(\gamma)-\alpha\z_t(\gamma)\\ \label{eq:dnt}
    \frac{d\n_t(\gamma)}{d\gamma}&=\bfP^\top\z_t(\gamma)-\n_t(\gamma),
\end{align}
where the components of $\z_t$ are represented by the spiking activity of pyramidal neurons, the components of $\n_t$ are represented by the activity of inhibitory interneurons, the components of $\bfP$ are represented by the synaptic weights from the interneurons to the pyramidal neurons, the components of $\bfP^\top$ are represented by the synaptic weights from the pyramidal neurons to the interneurons, and $\alpha$ is the threshold parameter of the adaptive algorithm.
These dynamics equilibrate at $\n_t=\bfP^\top\z_t$ and
\begin{align}
    \label{eq:zwhite}    
    \z_t&=(\bfP\bfP^\top+\alpha\I_k)^{-1}(\a_t+\b_t).
\end{align}
Provided $\alpha>0$, we can rearrange Equation~\eqref{eq:zwhite} to write the output as
\begin{align*}
    \z_t=\alpha^{-1}(\b_t+{\bf c}_t^a),
\end{align*}
where the components of ${\bf c}_t^a:=\a_t-\bfP\n_t$ are represented by the apical calcium plateau potentials within each pyramidal neuron.
In other words, the output is proportional to the sum of the basal dendritic current and the apical calcium plateau potential, which is consistent with experimental evidence showing that the output depends on both the basal inputs and apical calcium plateau potential \citep{bittner2015,bittner2017behavioral,GreinbergerMagee2020}.

Next, we compare the synaptic update rules with experimental evidence.
Rearranging Equation~\eqref{eq:zwhite} and substituting into the synaptic update rules in Algorithm~\ref{alg:adaptive}, we can rewrite the synaptic updates as follows:
\begin{align*}
    \W_x&\gets\W_x+\eta\left({\bf c}_t^b + (1-\alpha)\z_t\right)\x_t^\top\\
    \W_y&\gets\W_y+\eta\left({\bf c}_t^a + (1-\alpha)\z_t\right)\y_t^\top\\
    \bfP&\gets\bfP+\frac{\eta}{\vareps}(\z_t\n_t^\top-\bfP),
\end{align*}
where the components of ${\bf c}_t^b:=\b_t-\bfP\n_t$ are represented by basal calcium plateau potentials within each pyramidal neuron.
The learning signal for the basal (resp.\ apical) synaptic updates of this circuit is the correlation between the sum of the apical (resp.\ basal) calcium plateau potentials plus the scaled spiking activity of the pyramidal neurons, $\c_t^b+(1-\alpha)\z_t$ [resp.\ $\c_t^a+(1-\alpha)\z_t$], and the synaptic inputs to the basal (resp.\ apical) dendrites, $\x_t$ (resp.\ $\y_t$). When $\alpha=1$ the spiking (action potentials) of the post-synaptic neuron is not required for synaptic plasticity, whereas when $\alpha\ne1$ the spiking of the post-synaptic neuron affects synaptic plasticity along with the calcium plateau. Therefore, our model can account for a range of experimental observations which have demonstrated that spiking in the post-synaptic neuron contributes to plasticity in some contexts \citep{golding2002dendritic,gambino2014sensory,bittner2017behavioral,GreinbergerMagee2020}, but does not appear to affect plasticity in other contexts \citep{tigaret2016coordinated,sjostrom2006cooperative}.
Unlike Hebbian learning rules which depend only on the correlation of the spiking output of the post-synaptic neuron with the pre-synaptic spiking, the mechanisms involving the calcium plateau potential represented internally in a neuron are called non-Hebbian. 
Because synapses have access to both the corresponding presynaptic activity and to the calcium plateau potential, the learning rule remains local.

Note that the update rule for the synapses in the apical dendrites, $\W_x$, depend on the basal calcium plateau potentials $\c_t^b$.
Experimental evidence is focused on apical calcium plateau potentials and it is not clear whether differences between basal inputs and inhibitory signals generate calcium signals for driving plasticity in the apical dendrites.
Alternatively, the learning rule for $\W_x$ coincides with the learning rule for the apical dendrites in \citep{golkar2020}, where a biological implementation in terms of local depolarization and backpropagating spikes was proposed.
Due to the inconclusive evidence pertaining to plasticity in the apical tuft, we find it useful to put forth both interpretations.

Multi-compartmental models of pyramidal neurons have been invoked previously in the context of biological implementation of the backpropagation algorithm \citep{kording2001supervised,urbanczik2014learning,guerguiev2017towards,haga2018dendritic,sacramento2018dendritic,richards2019dendritic}. 
Under this interpretation, the apical compartment represents the target output, the basal compartment represents the algorithm prediction and calcium plateau potentials communicate the error from the apical to the basal compartment, which is used for synaptic weight updates. 
The difference between these models and ours is that we use a normative approach to derive not only the learning rules but also the neural dynamics of the CCA algorithm ensuring that the output of the network is known for any input. 
On the other hand, the linearity of neural dynamics in our network means that stacking our networks will not lead to any nontrivial results expected of a deep learning architecture. We leave introducing nonlinearities into neural dynamics and stacking our network to future work.

We conclude this section with comments on the interneuron-to-pyramidal neuron synaptic weight matrix $\bfP$ and pyramidal neuron-to-interneuron synaptic weight matrix $\bfP^\top$, as well as the computational role of the interneurons in this network. First, the algorithm appears to require a weight sharing mechanism between the two sets of synapses to ensure the symmetry between the weight matrices, which is biologically unrealistic and commonly referred to as the weight transport problem. However, even without any initial symmetry between these feedforward and feedback synaptic weights, because of the symmetry of the local learning rules, the difference between the two will decay exponentially without requiring weight transport (see Appendix \ref{apdx:decouple}). Second, in Equations \eqref{eq:dzt}--\eqref{eq:dnt}, the interneuron-to-pyramidal neuron synaptic weight matrix $\bfP$ is preceded by a negative sign and the pyramidal neuron-to-interneuron synaptic weight matrix $\bfP^\top$ is preceded by a positive sign, which is consistent with the fact that, in simplified cortical microcircuits, interneuron-to-pyramidal neuron synapses are inhibitory whereas the pyramidal neuron-to-interneuron synapses are excitatory. That being said, this interpretation is superficial because the weight matrices are not constrained to be non-negative, which is due to the fact that we are implementing a linear statistical method. 
Imposing non-negativity constraints on the weights $\bfP$ and $\bfP^\top$ may be useful for implementing nonlinear statistical methods; however, this requires further investigation.
Finally, the activities of the interneurons $\N$ were introduced in Equation~\eqref{eq:whitening} to decorrelate the output. This is consistent with previous models of the cortex (e.g., \cite{king2013inhibitory,wanner2020whitening}), which have introduced inhibitory interneurons to decorrelate excitatory outputs; however, in contrast to the current work, the models proposed in \cite{king2013inhibitory,wanner2020whitening} are not normative.

\section{Numerical experiments}
\label{sec:numerics}

We now evaluate the performance of the online algorithms, Bio-CCA and Adaptive Bio-CCA with output whitening.
In each plot, the lines and shaded regions respectively denote the means and 90\% confidence intervals over 5 runs.
Detailed descriptions of the implementations are given in Appendix~\ref{apdx:details}.
All experiments were performed in Python on an iMac Pro equipped with a 3.2 GHz 8-Core Intel Xeon W CPU.
The evaluation code is available at \url{https://github.com/flatironinstitute/bio-cca}.

\subsection{Datasets}
\label{sec:datasets}

We first describe the evaluation datasets.

\begin{figure}[ht]
    \centering
    \includegraphics[width=.95\textwidth]{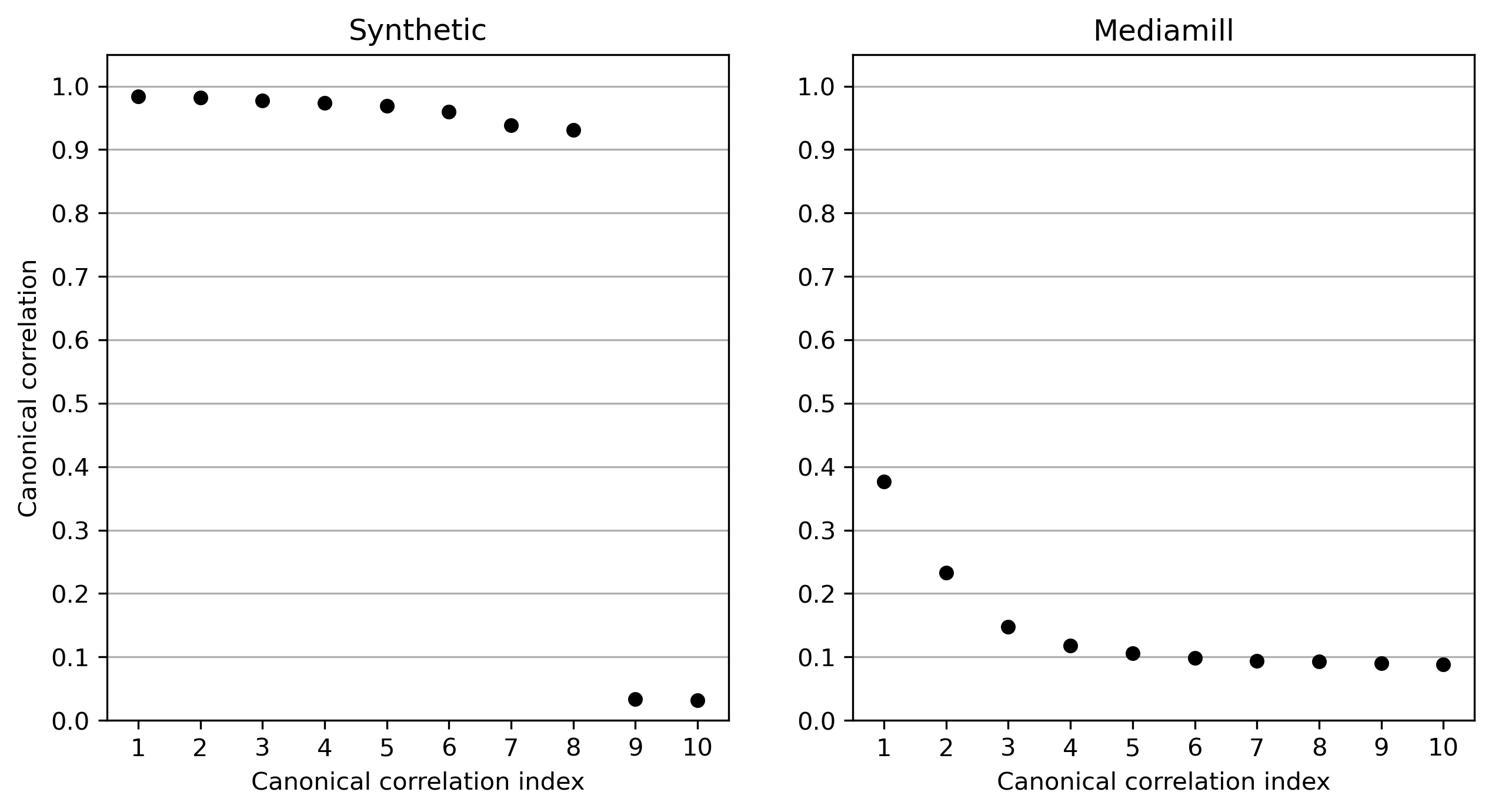}
    \caption{Top 10 canonical correlations $\rho_1,\dots,\rho_{10}$ of the synthetic dataset (left) and \texttt{Mediamill} (right).} \label{fig:can_cor}
\end{figure}

\paragraph{Synthetic.} We generated a synthetic dataset with $T=100,000$ samples according to the probabilistic model for CCA introduced by \citet{bach2005probabilistic}. 
In particular, let ${\bf s}_1,\dots,{\bf s}_T$ be i.i.d.\ 8-dimensional latent mean-zero Gaussian vectors with identity covariance.
Let ${\bf T}_x\in\R^{50\times 8}$, ${\bf T}_y\in\R^{30\times 8}$, $\boldsymbol{\Psi}_x\in\mathcal{S}_{++}^{50}$ and $\boldsymbol{\Psi}_y\in\mathcal{S}_{++}^{30}$ be randomly generated matrices and define the 50-dimensional observations $\x_1,\dots,\x_T$ and 30-dimensional observations $\y_1,\dots,\y_T$ by
\begin{align*}
    \x_t:={\bf T}_x{\bf s}_t+\boldsymbol{\phi}_t,&&\y_t:={\bf T}_y{\bf s}_t+\boldsymbol{\psi}_t,&&t=1,\dots,T,
\end{align*}
where $\boldsymbol{\phi}_1,\dots,\boldsymbol{\phi}_T$ (resp.\ $\boldsymbol{\psi}_1,\dots,\boldsymbol{\psi}_T$) are i.i.d.\ 50-dimensional (resp.\ 30-dimensional) mean-zero Gaussian vectors with covariance $\boldsymbol{\Psi}_x$ (resp.\ $\boldsymbol{\Psi}_y$).
Thus, conditioned on the latent random variable ${\bf s}$, the observation $\x$ (resp.\ $\y$) has a Gaussian distribution with mean ${\bf T}_x{\bf s}$ (resp.\ ${\bf T}_y{\bf s}$) with covariance $\boldsymbol{\Psi}_x$ (resp.\ $\boldsymbol{\Psi}_y$), i.e.,
\begin{align*}
    \x_t|{\bf s}_t\sim\mathcal{N}({\bf T}_x{\bf s}_t,\boldsymbol{\Psi}_x),&&\y_t|{\bf s}_t\sim\mathcal{N}({\bf T}_y{\bf s}_t,\boldsymbol{\Psi}_y).
\end{align*}
For this generative model, \citet{bach2005probabilistic} showed that the posterior expectation of the latent vector ${\bf s}_t$ given the observation $(\x_t,\y_t)$ is a linear transformation of the sum of the $8$-dimensional CCSPs $\z_t$; that is, $\mathbb{E}\left[{\bf s}_t|(\x_t,\y_t)\right]={\bf L}{\bf z}_t$ for some $8\times 8$ matrix ${\bf L}$. (To see this, set $M_1=M_2=P_d^{1/2}$ in the paragraph following \cite[Theorem 2]{bach2005probabilistic}.)
The first 10 canonical correlations are plotted in Figure \ref{fig:can_cor} (left).
Observe that the first 8 canonical correlations are close to 1 and the remaining canonical correlations are approximately 0.
This sharp drop in the canonical correlations is a consequence of the linear generative model and is generally not the case in real data (see, e.g., the right panel in Figure \ref{fig:can_cor}).
Still, we find it useful to test our algorithms on this synthetic dataset since the generative model is well studied and relevant to CCA \citep{bach2005probabilistic}.

\paragraph{Mediamill.}

The dataset \texttt{Mediamill} \citep{snoek2006challenge} consists of $T=43,907$ samples (including training and testing sets) of video data and text annotations, and has been previously used to evaluate CCA algorithms \citep{arora2017stochastic,Zhao2020}. 
The first view consists of 120-dimensional visual features extracted from representative video frames.
The second view consists of 101-dimensional vectors whose components correspond to manually labeled semantic concepts associated with the video frames (e.g., ``basketball'' or ``tree'').
To ensure that the problem is well-conditioned, we add Gaussian noise with covariance matrix $\varepsilon \I_{120} $ (resp.\ $\varepsilon \I_{101}$), for $\varepsilon= 0.1$, to the first (resp.\ second) view to generate the data matrix $\X$ (resp.\ $\Y$).
The first 10 canonical correlations are plotted in Figure \ref{fig:can_cor} (right).

\paragraph{Non-stationary.} To evaluate Adaptive Bio-CCA with output whitening, we generated a non-stationary synthetic dataset with $T=300,000$ samples, which are streamed from 3 distinct distributions that are generated according to the probabilistic model in \citep{bach2005probabilistic}. In this case, the first $N=100,000$ samples are generated from a 4-dimensional latent source, the second $N$ samples are generated from an 8-dimensional latent source, and the final $N$ samples are generated from a 1-dimensional latent source.

\begin{figure}[ht]
    \centering
    \includegraphics[width=.95\textwidth]{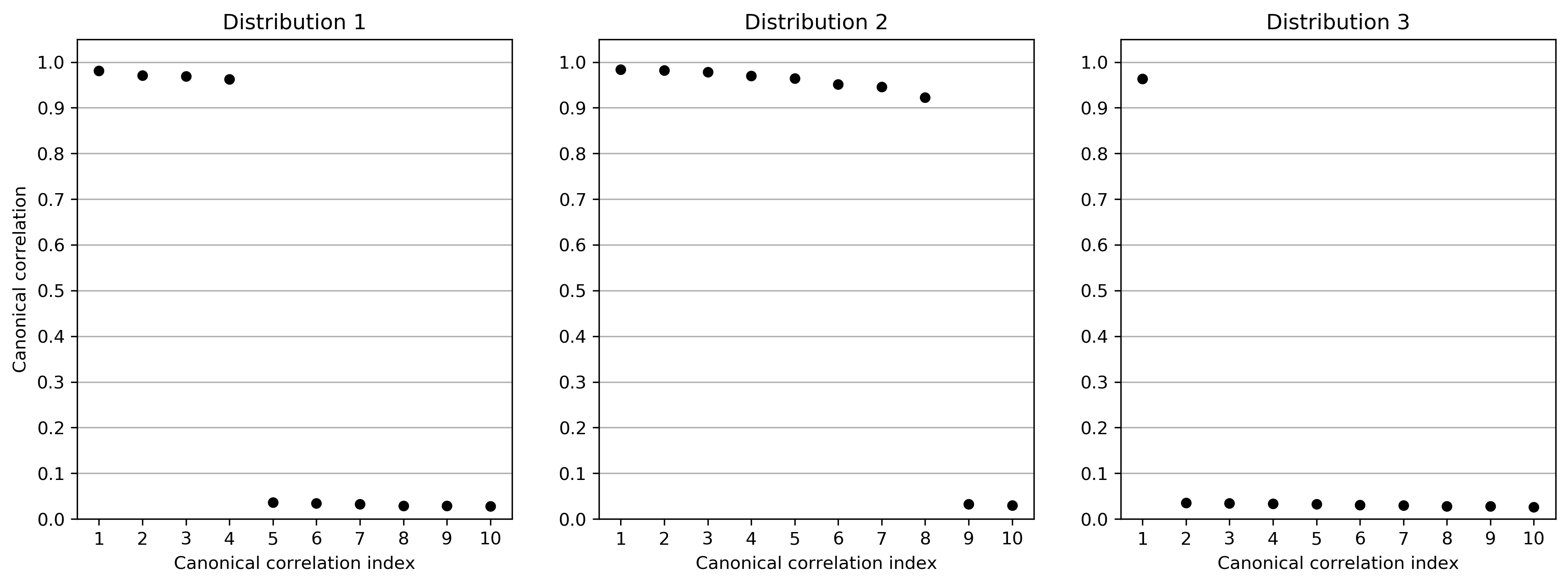}
    \caption{Top 10 canonical correlations $\rho_1,\dots,\rho_{10}$ of the 3 distributions that contribute to the non-stationary synthetic dataset.} \label{fig:adaptive_can_cor}
\end{figure}

Specifically, we let ${\bf s}_1,\dots,{\bf s}_{N}$ (resp.\ ${\bf s}_{N+1},\dots,{\bf s}_{2N}$ and ${\bf s}_{2N+1},\dots,{\bf s}_{T}$) be i.i.d.\ 4-dimensional (resp.\ 8-dimensional and 1-dimensional) mean-zero Gaussian vectors with identity covariance. We then let ${\bf T}_x^{(1)}\in\R^{50\times 4}$, ${\bf T}_x^{(2)}\in\R^{50\times 8}$, ${\bf T}_x^{(3)}\in\R^{50\times 1}$, ${\bf T}_y^{(1)}\in\R^{30\times 4}$, ${\bf T}_y^{(2)}\in\R^{30\times 8}$, ${\bf T}_y^{(3)}\in\R^{30\times 1}$, $\boldsymbol{\Psi}_x\in\mathcal{S}_{++}^{50}$ and $\boldsymbol{\Psi}_y\in\mathcal{S}_{++}^{30}$ be randomly generated matrices and define the 50-dimensional observations $\x_1,\dots,\x_T$ and 30-dimensional observations $\y_1,\dots,\y_T$ by
\begin{align*}
    \x_t:={\bf T}_x^{(1)}{\bf s}_t+\boldsymbol{\phi}_t,&&\y_t:={\bf T}_y^{(1)}{\bf s}_t+\boldsymbol{\psi}_t,&&t=1,\dots,N,\\
    \x_t:={\bf T}_x^{(2)}{\bf s}_t+\boldsymbol{\phi}_t,&&\y_t:={\bf T}_y^{(2)}{\bf    s}_t+\boldsymbol{\psi}_t,&&t=N+1,\dots,2N,\\
    \x_t:={\bf T}_x^{(3)}{\bf s}_t+\boldsymbol{\phi}_t,&&\y_t:={\bf T}_y^{(3)}{\bf s}_t+\boldsymbol{\psi}_t,&&t=2N+1,\dots,T,
\end{align*}
where, as before, $\boldsymbol{\phi}_1,\dots,\boldsymbol{\phi}_T$ (resp.\ $\boldsymbol{\psi}_1,\dots,\boldsymbol{\psi}_T$) are i.i.d.\ 50-dimensional (resp.\ 30-dimensional) mean-zero Gaussian vectors with covariance $\boldsymbol{\Psi}_x$ (resp.\ $\boldsymbol{\Psi}_y$).
In Figure \ref{fig:adaptive_can_cor}, we plot the first 10 canonical correlations for each of the 3 distributions.

\subsection{Bio-CCA}

We now evaluate the performance of Bio-CCA (Algorithm \ref{alg:online}) on the synthetic dataset and \texttt{Mediamill}.

\paragraph{Competing algorithms.} We compare the performance of Bio-CCA with the following state-of-the-art online CCA algorithms: 
\begin{itemize}
    \item A two-time-scale algorithm for computing the top canonical correlation basis vectors (i.e., $k=1$) introduced by \citet{bhatia2018gen}. The algorithm is abbreviated ``Gen-Oja'' due to its resemblance to Oja's method \citep{oja1982simplified}.
    \item An inexact Matrix Stochastic Gradient method for solving CCA, abbreviated ``MSG-CCA'', which was derived by \citet{arora2017stochastic}.
    \item The asymmetric neural network proposed by \citet{Zhao2020}, which we abbreviate as ``Asym-NN''.
    \item The biologically plausible Reduced-Rank Regression algorithm derived by \citet{golkar2020}, abbreviated ``Bio-RRR'', which implements a supervised version of CCA when $s=1$ (see Algorithm \ref{alg:rrr} in Appendix \ref{apdx:compare}).
\end{itemize}
Detailed descriptions of the implementations of each algorithm are provided in Appendix \ref{apdx:details}.

\paragraph{Performance metrics.} To evaluate the performance of Bio-CCA, we use 2 performance metrics.
The first performance metric is the following $[0,2]$-valued normalized objective error function:
\begin{align}\label{eq:obj}
    \text{Normalized objective error}(t):=\frac{\rho_\text{max}-\tr(\V_{x,t}^\top\Sig_{xy}\V_{y,t})}{\rho_\text{max}}.
\end{align}
Here $\rho_\text{max}:=(\rho_1+\cdots+\rho_k)/2$ is the optimal value of the CCA objective \eqref{eq:cca1}--\eqref{eq:cca2}, and $(\V_{x,t},\V_{y,t})$ are the basis vectors reported by the respective algorithm after iteration $t$, normalized to ensure they satisfy the orthonormality constraint \eqref{eq:cca2}.
(We do not evaluate Bio-RRR using this metric because the algorithm only outputs one set of basis vectors.)

The second performance metric is the ($x-$)subspace error function is defined by
\begin{align}\label{eq:subspace_error}
    \text{Subspace error}(t)&:=\norm{\V_{x,t}(\V_{x,t}^\top\V_{x,t})^{-1}\V_{x,t}^\top-\ol\V_x(\ol\V_x^\top\ol\V_x)^{-1}\ol\V_x^\top}_\text{F}^2,
\end{align}
where $\ol\V_x$ is the matrix of optimal basis vectors defined as in Equation \eqref{eq:Vbar}.
(We do not evaluate MSG-CCA using this metric because the algorithm outputs the product $\V_{x,t}\V_{y,t}^\top$ rather than outputting the basis vectors $\V_{x,t}$ and $\V_{y,t}$ separately.)

\begin{figure}
    \centering
    \includegraphics[width=.95\textwidth]{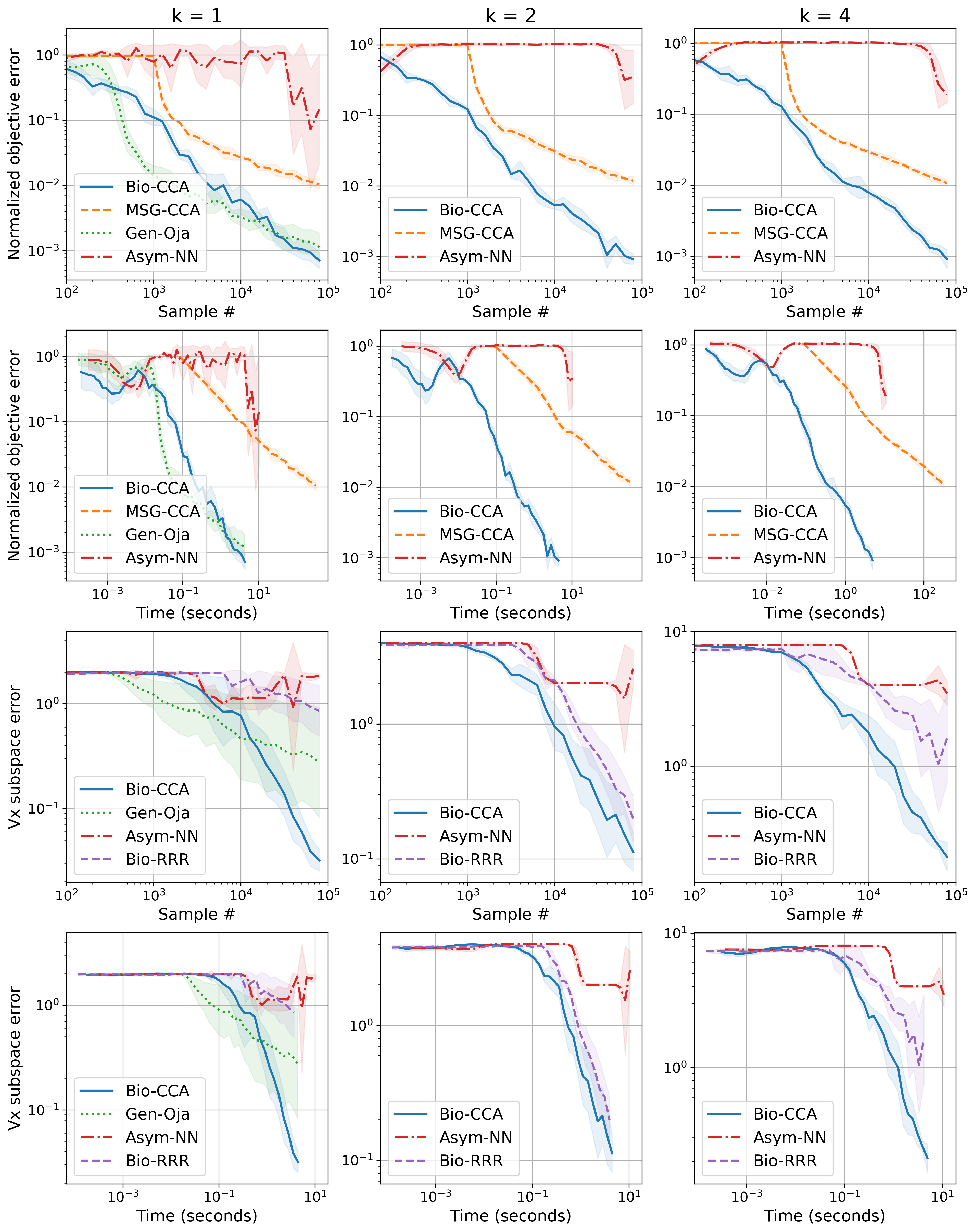}
    \caption{Comparisons of Bio-CCA (Algorithm \ref{alg:online}) with the competing algorithms on the synthetic dataset, for $k=1,2,4$, in terms of the normalized objective error defined in Equation~\eqref{eq:obj} as a function of sample number and runtime (top two rows), and in terms of the subspace error defined in Equation~\eqref{eq:subspace_error} as a function of sample number and runtime (bottom two rows).} \label{fig:biocca_synthetic_error}
\end{figure}

\paragraph{Evaluation on the synthetic dataset.} In Figure~\ref{fig:biocca_synthetic_error} we plot the performance of Bio-CCA, in terms of both sample and runtime efficiency, against the competing algorithms for target dimensions $k=1,2,4$ on the synthetic dataset, presented once in a randomly permuted order.
For $k=1$, Gen-Oja initially outperforms Bio-CCA in sample and runtime efficiency; however Bio-CCA eventually performs comparably with Gen-Oja when given sufficiently many samples (and outperforms Gen-Oja in terms of subspace error).
For $k=2,4$, the sample and runtime efficiency of Bio-CCA outperforms the other competing algorithms.
The MSG-CCA error does not begin to decay until the $10^3$ iteration because the first $10^3$ samples are used to obtain initial estimates of the covariance matrices $\Sig_{xx}$ and $\Sig_{yy}$.
The poor performance of the asymmetric network \citep{Zhao2020} is due in part to the fact that the algorithm depends on the gaps between the canonical correlations, which are small for the synthetic dataset.
In Figure \ref{fig:ortho1} of Appendix \ref{apdx:ortho}, we verify that the Bio-CCA basis vectors asymptotically satisfy the orthonormality constraint \eqref{eq:cca2}.

\begin{figure}
    \centering
    \includegraphics[width=.95\textwidth]{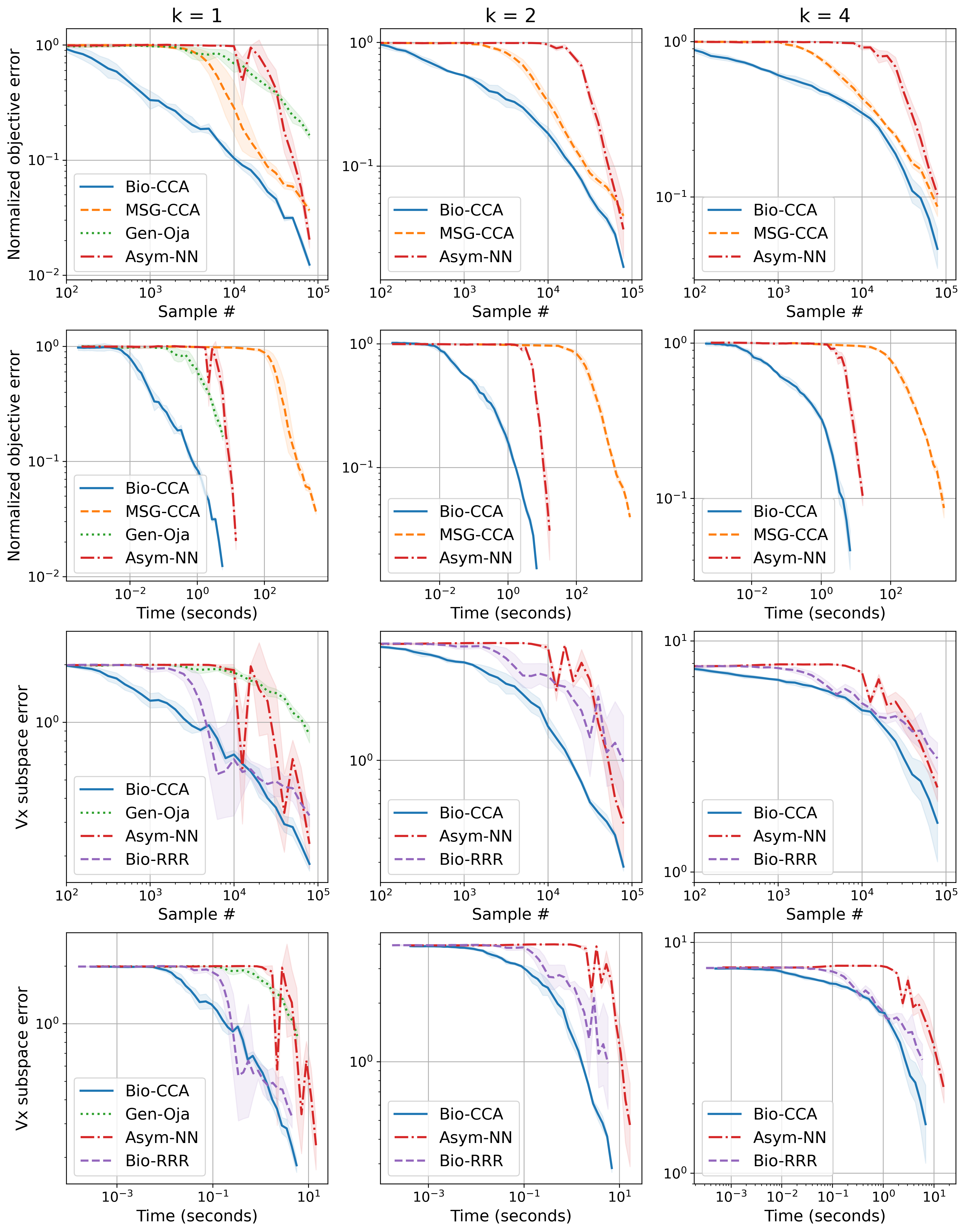}
    \caption{Comparisons of Bio-CCA (Algorithm \ref{alg:online}) with the competing algorithms on \texttt{Mediamill}, for $k=1,2,4$, in terms of the normalized objective error defined in Equation~\eqref{eq:obj} as a function of sample number and runtime (top two rows), and in terms of the subspace error defined in Equation~\eqref{eq:subspace_error} as a function of sample number and runtime (bottom two rows).} \label{fig:biocca_mediamill_error}
\end{figure}

\paragraph{Evaluation on Mediamill.} In Figure~\ref{fig:biocca_mediamill_error} we plot the performance of Bio-CCA, in terms of both sample and runtime efficiency, against the competing algorithms for target dimensions $k=1,2,4$ on \texttt{Mediamill}, presented 3 times with a randomly permuted order in each presentation.
When tested on \texttt{Mediamill}, the sample and runtime efficiency of Bio-CCA outperform the competing algorithms (for $k=1$, Bio-RRR performs comparably with Bio-CCA).
In Figure \ref{fig:ortho2} of Appendix \ref{apdx:ortho}, we verify that the Bio-CCA basis vectors asymptotically satisfy the orthonormality constraint \eqref{eq:cca2}.

\subsection{Adaptive Bio-CCA with output whitening}

Next, we evaluate the performance of Adaptive Bio-CCA with output whitening (Algorithm \ref{alg:adaptive}) on all 3 datasets.
Since we are unaware of competing online algorithms for adaptive CCA with output whitening, to compare the performance of Algorithm \ref{alg:adaptive} to existing methods, we also plot the performance of Bio-RRR \cite{golkar2020} with respect to subspace error, where we a priori select the target dimension. 
We chose Bio-RRR because the algorithm also maps onto a neural network that resembles the cortical microcircuit and because it performs relatively well on the synthetic dataset and \texttt{Mediamill}.

\paragraph{Performance metric.} To evaluate the performance of Adaptive Bio-CCA with output whitening, we use a subspace error metric.
Recall that the target output rank of Algorithm \ref{alg:adaptive}, denoted $r$, is equal to the number of canonical correlations $\rho_1,\dots,\rho_k$ that exceed $\max(\alpha-1,0)$, i.e.,
\begin{align}
    \label{eq:r}
    r:=\max\{1\le i\le k:\rho_i>\max(\alpha-1,0)\}.
\end{align}
Since the target dimension is often less than $k$, the subspace error defined in Equation \eqref{eq:subspace_error} is not an appropriate metric because the projection matrices are rank $k$.
Rather, we evaluate Adaptive Bio-CCA with output whitening using the \textit{adaptive} subspace error function defined by
\begin{align}\label{eq:adaptive_subspace_error}
    \text{Adaptive subspace error}(t)&:=\norm{\widetilde\U_{x,t}\widetilde\U_{x,t}^\top-\widetilde\V_x(\widetilde\V_x^\top\widetilde\V_x)^{-1}\widetilde\V_x^\top}_\text{F}^2,
\end{align}
where $\widetilde\U_{x,t}$ is the $m\times r$ matrix whose column vectors are the top $r$ right-singular vectors of $\W_{x,t}$, and $\widetilde\V_x$ is the $m\times r$ matrix whose $i$\textsuperscript{th} column vector is equal to the $i$\textsuperscript{th} column vector of the matrix $\ol\V_x$ defined in Equation \eqref{eq:Vbar}, for $i=1,\dots,r$.
(The covariance matrices used in the definition for $\ol\V_x$ are those associated with the distribution that is being streamed at iteration $t$.)
If $r=k$, then the error aligns with the subspace error defined in Equation \eqref{eq:subspace_error}.
Since the output dimension of Bio-RRR is a priori set to $r$, the performance of Bio-RRR is evaluated using the subspace error defined in Equation \eqref{eq:subspace_error}.

\begin{figure}
    \centering
    \includegraphics[width=.95\textwidth]{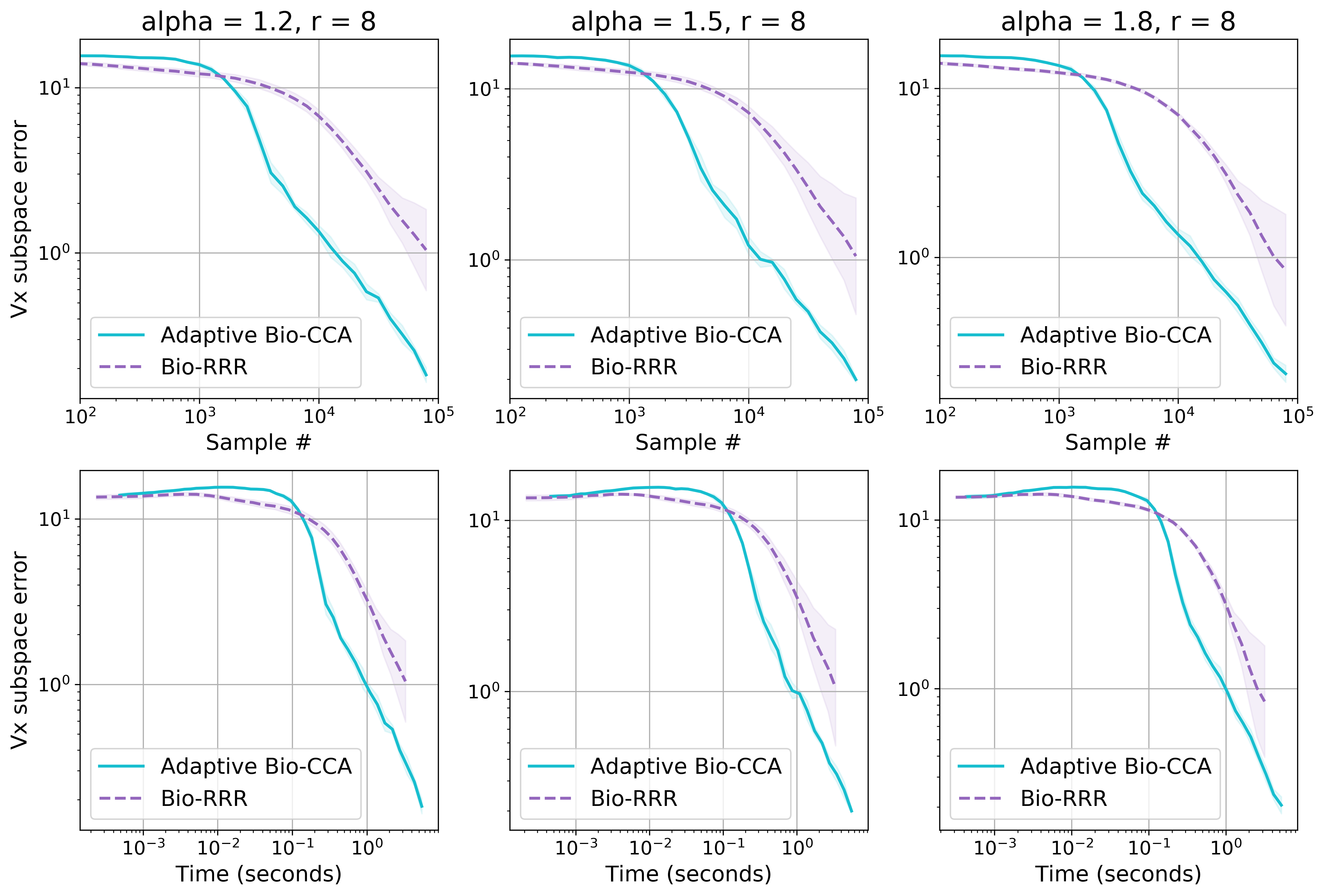}
    \caption{Comparisons of Adaptive Bio-CCA with output whitening (Algorithm \ref{alg:adaptive}) and Bio-RRR on the synthetic dataset, for $\alpha=1.2,1.5,1.8$, in terms of the subspace error as a function of sample number (top row) and runtime (bottom row). For each $\alpha$, $r$ is the target output rank defined as in Equation \eqref{eq:r}.} \label{fig:adaptive_synth}
\end{figure}

\paragraph{Evaluation on the synthetic dataset.} From Figure \ref{fig:can_cor} (left) we see that the first 8 canonical correlations are close to 1 and the remaining canonical correlations are approximately 0.
Therefore, for $k\ge8$ and $\alpha\in(1.1,1.9)$, Algorithm \ref{alg:adaptive} should project the inputs $\x_t$ and $\y_t$ onto the 8-dimensional subspace spanned by the top 8 pairs of canonical correlation basis vectors, and set the non-zero output covariance eigenvalues to one.
In Figure \ref{fig:adaptive_synth} we plot the performance of Algorithm \ref{alg:adaptive} with $k=10$ for $\alpha=1.2,1.5,1.8$ on the synthetic dataset (presented once with a randomly permuted order).
We see that Adaptive Bio-CCA with output whitening outperforms Bio-RRR, even though the target output dimension of Bio-RRR was set a priori. 
In Figure \ref{fig:white1} of Appendix \ref{apdx:ortho}, we verify that the Adaptive Bio-CCA with output whitening basis vectors asymptotically satisfy the whitenening constraint.

\begin{figure}
    \centering
    \includegraphics[width=.95\textwidth]{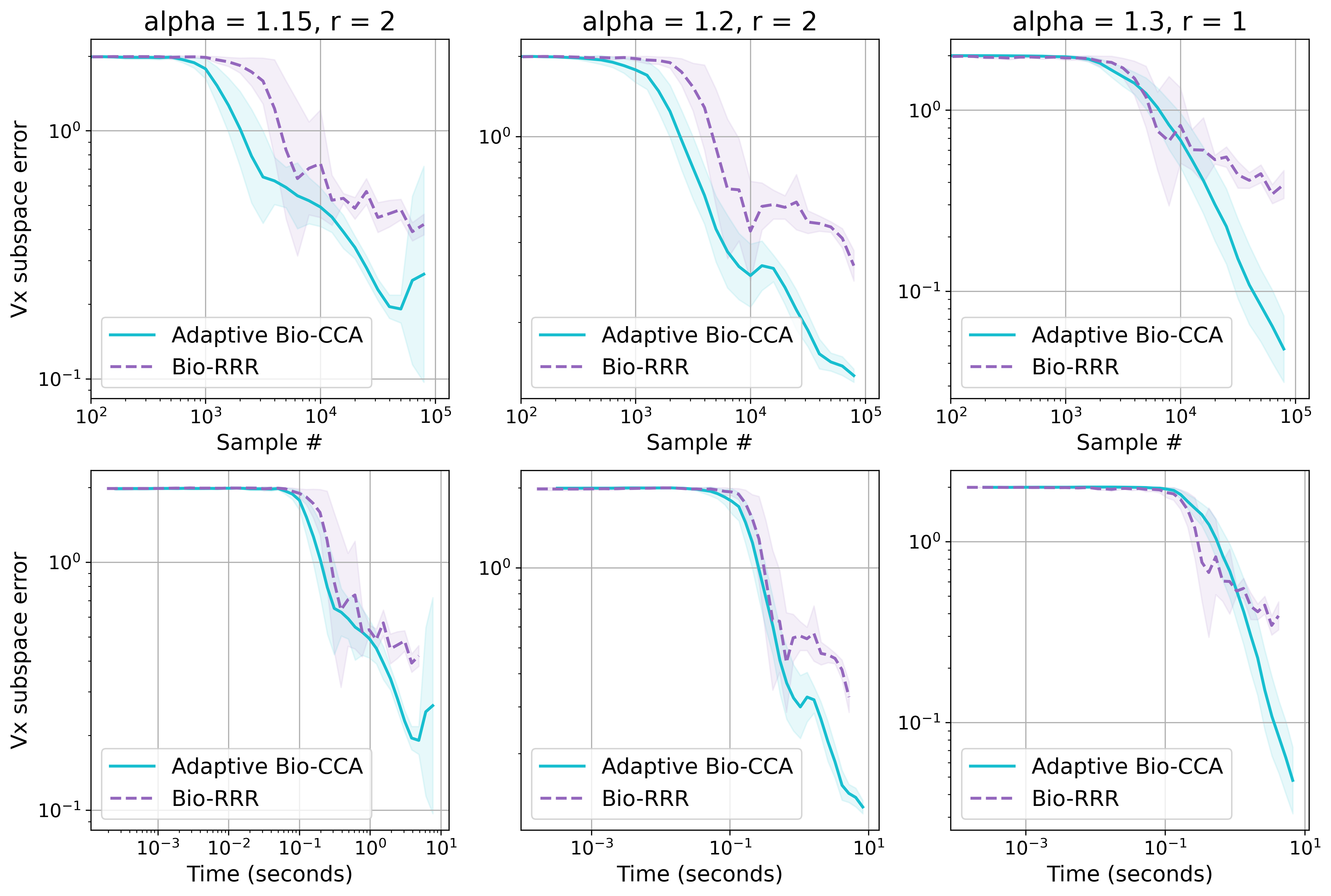}
    \caption{Comparisons of Adaptive Bio-CCA with output whitening (Algorithm \ref{alg:adaptive}) with Bio-RRR on the dataset \texttt{Mediamill}, for $\alpha=1.2,1.5,1.8$, in terms of the subspace error as a function of sample number (top row) and runtime (bottom row). For each $\alpha$, $r$ is the target output rank defined as in Equation \eqref{eq:r}.} \label{fig:adaptive_mm}
\end{figure}

\paragraph{Evaluation on Mediamill.} From Figure \ref{fig:can_cor} (right) we see that the canonical correlations of the dataset \texttt{Mediamill} exhibit a more gradual decay than the canonical correlations of the synthetic dataset.
As we increase the threshold $\alpha$ in the interval $(1.1,1.4)$, the rank of the output of Algorithm \ref{alg:adaptive} decreases.
In Figure \ref{fig:adaptive_mm} we plot the performance of Algorithm \ref{alg:adaptive} with $k=10$ for $\alpha=1.15,1.2,1.3$ on \texttt{Mediamill} (presented three times with a randomly permuted order in each presentation).
As with the synthetic dataset, we find that Adaptive Bio-CCA with output whitening outperforms Bio-RRR.
In Figure \ref{fig:white2} of Appendix \ref{apdx:ortho}, we verify that the Adaptive Bio-CCA with output whitening basis vectors asymptotically satisfy the whitenening constraint.

\begin{figure}
    \centering
    \includegraphics[width=.95\textwidth]{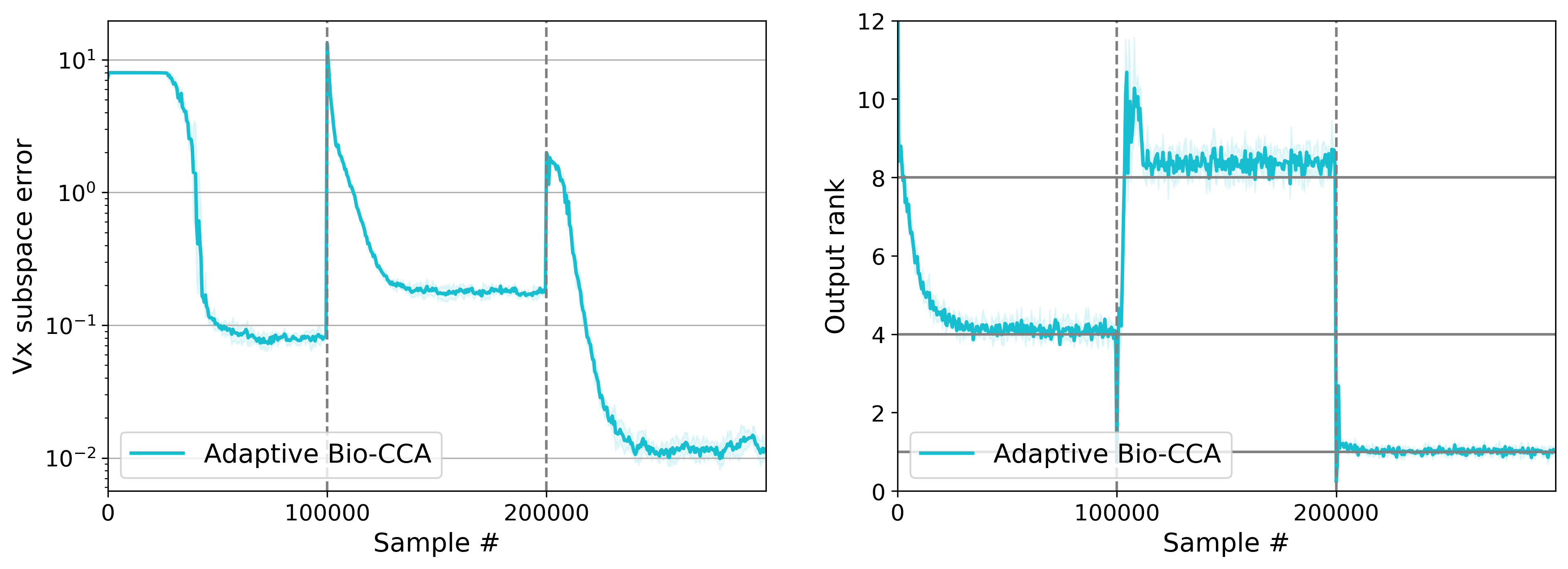}
    \caption{On the left we plot the performance of Adaptive Bio-CCA with output whitening (with $\alpha=1.5$) on the non-stationary synthetic dataset in terms of the subspace error defined in Equation~\eqref{eq:adaptive_subspace_error}.
    On the right we plot the output rank defined in Equation \eqref{eq:output_rank}.}
    \label{fig:nonstationary}
\end{figure}

\paragraph{Evaluation on the non-stationary dataset.} From Figure \ref{fig:adaptive_can_cor}, we see that the distributions that contribute to the non-stationary dataset all have canonical correlations that are close to 0 or 1. The first distribution has 4 canonical correlations close to 1, the second distribution has 8 canonical correlations close to 1 and the third distribution has 1 canonical correlation close to 1.
Therefore, for $k\ge8$ and $\alpha\in(1.1,1.9)$, Algorithm \ref{alg:adaptive} should initially (for the first 100,000 inputs) project the inputs $\x_t$ and $\y_t$ onto a 4-dimensional subspaces, then (for the second 100,000 inputs) project the inputs onto an 8-dimensional subspaces, and finally (for the final 100,000 inputs) project the inputs onto a 1-dimensional subspaces.
In Figure \ref{fig:nonstationary} (left) we plot the performance of Algorithm \ref{alg:adaptive} with $k=10$ and $\alpha=1.5$ on the non-stationary dataset in terms of the adaptive subspace error.\footnote{Since the competing algorithms are not adaptive and need to have their output dimension set by hand, we do not include a competing algorithm for comparison.}
Note that the adaptive subspace error spikes each time the algorithm is streamed inputs from a new distribution (these epochs are denoted by the gray vertical dashed lines) before decaying. In Figure \ref{fig:nonstationary} (right) we plot the output rank of Adaptive Bio-CCA with output whitening defined by
\begin{align}
    \label{eq:output_rank}
    \text{Output rank}(t)=\tr(\V_{x,t}^\top\Sig_{xx}\V_{x,t}+\V_{x,t}^\top\Sig_{xy}\V_{y,t}+\V_{y,t}^\top\Sig_{xy}^\top\V_{x,t}+\V_{y,t}^\top\Sig_{yy}\V_{y,t}),
\end{align}
where, at each iteration $t$, the covariance matrices correspond to the distribution from which $(\x_t,\y_t)$ is streamed.
Note that the output rank adapts each time the algorithm is streamed inputs from a new distribution to match the dimension of the latent variable generating the distribution.
In particular, we see that the algorithm is able to quickly adapt its output rank to new distributions.

\section{Discussion}

In this work, we derived an online algorithm for CCA that can be implemented in a neural network with multi-compartmental neurons and local, non-Hebbian learning rules.
We also derived an extension that adaptively chooses the output rank and whitens the output.
Remarkably, the neural architecture and non-Hebbian learning rules of our extension resembled neural circuitry and non-Hebbian plasticity in cortical pyramidal neurons.
Thus, our neural network model may be useful for understanding the computational role of multi-compartmental neurons with non-Hebbian plasticity.

While our neural network model captures salient features of cortical microcircuits, there are important biophysical properties that are not explained by our model.
First, our model uses linear neurons to solve the linear CCA problem, which substantially limits its computational capabilities and is a major simplification of cortical pyramidal neurons which can perform nonlinear operations \citep{gidon2020dendritic}.
However, studying the analytically tractable and interpretable linear neural network model is useful for understanding more complex nonlinear models.
Such an approach has proven successful for studying deep networks in the machine learning literature \citep{arora2018a}.
In future work, we plan to incorporate nonlinear neurons in our model.

Second, our neural network implementation requires the same number of interneurons and principal neurons, whereas in the cortex there are approximately 4 times more pyramidal neurons than interneurons \citep{larkum1999new}.
In our model, the interneurons decorrelate the output and, in practice, the optimal fixed points of the algorithm can destabilize when there are fewer interneurons than principal neurons (see Remark 3A in the supplementary material of \citep{pehlevan2015normative}). 
In biological circuits, these instabilities could be mitigated by other biophysical constraints; however, a theoretical justification would require additional work.

Third, the output of our neural network is the equally weighted sum of the basal and apical projections.
However, experimental evidence suggests that the pyramidal neurons integrate their apical and basal inputs asymmetrically \citep{larkum2009synaptic,Larkum2013,major2013active}. 
In addition, in our model, the apical learning rule is non-Hebbian and depends on a calcium plateau potential that travels from the basal dendrites to the apical tuft.
Experimental evidence for calcium plateau potential dependent plasticity is focused on the basal dendrites, with inconclusive evidence on the plasticity rules for the apical dendrites \citep{golding2002dendritic,sjostrom2006cooperative}.

To provide an alternative explanation of cortical computation, in a separate work \citep{golkar2020}, we derive an online algorithm for the general \textit{supervised} learning method Reduced-Rank Regression, which includes CCA as a special case (see Appendix \ref{apdx:compare} for a detailed comparison of the two algorithms). 
The algorithm also maps onto a neural network with multi-compartmental neurons and non-Hebbian plasticity in the basal dendrites. 
Both models adopt a normative approach in which the algorithms are derived from principled objective functions.
This approach is highly instructive as the differences between the models highlight which features of the network that are central to implementing an unsupervised learning method versus a supervised learning method.

There are three main differences between the biological interpretation of the two algorithms.
First, the output of the network in \citep{golkar2020} is the projection of the basal inputs, with no apical contribution.
Second, the network in \citep{golkar2020} allows for a range of apical synaptic update rules, including Hebbian updates.
Third, the adaptive network derived here includes a threshold parameter $\alpha$, which adaptively sets the output dimension and is not included in \citep{golkar2020}.
In our model, this parameter corresponds the contribution of the somatic output to plasticity in the basal dendrites.
These differences can be compared to experimental outcomes to provide evidence that cortical microcircuits implement unsupervised algorithms, supervised algorithms, or mixtures of both.
Thus, we find it informative to put forth and contrast the two models.

Finally, we did not prove theoretical guarantees that our algorithms converge.
As we show in Appendix \ref{apdx:gda}, Offline-CCA and Bio-CCA can be viewed as gradient descent-ascent and stochastic gradient descent-ascent algorithms for solving a nonconvex-concave min-max problem.
While gradient descent-ascent algorithms are natural methods for solving such min-max problems, they are not always guaranteed to converge to a desired solution.
In fact, when the gradient descent step size not sufficiently small relative to the gradient ascent step size (i.e., when $\tau$ is not sufficiently small), gradient descent-ascent algorithms for solving nonconvex-concave min-max problems can converge to limit cycles \citep{hommes2012multiple,mertikopoulos2018cycles}.
Establishing local or global convergence, and convergence rate guarantees for general gradient descent-ascent algorithms is an active area of research, and even recent advances \citep{lin2019gradient} impose assumptions that are not satisfied in our setting.
In Appendix \ref{apdx:gda}, we discuss these challenges and place our algorithms within the broader context of gradient descent-ascent algorithms for solving nonconvex-concave min-max problems.

\section*{Acknowledgements}

We thank Nati Srebro for drawing our attention to CCA, and we thank Tiberiu Tesileanu and Charles Windolf for their helpful feedback on an earlier draft of this manuscript.

\appendix

\section{Sums of CCSPs as principal subspace projections}
\label{apdx:sm}

In this section, we show that the sums of the CCPSs $\z_1,\dots,\z_T$ can be expressed as the principal subspace projection of the whitened concatenated data $\boldsymbol{\xi}_1,\dots,\boldsymbol{\xi}_T$, defined by
\begin{align}
\label{eq:xi_apdx}
    \boldsymbol{\xi}_t:=
    \begin{bmatrix}
        \Sig_{xx}^{-1/2}\x_t \\
        \Sig_{yy}^{-1/2}\y_t
    \end{bmatrix}.
\end{align}
To this end, recall the CCA objective
\begin{align*}
    \argmax{\V_x\in\R^{m\times k},\V_y\in\R^{n\times k}}\tr\left(\V_x^\top\Sig_{xy}\V_y\right)
\end{align*}
subject to the whitening constraint $\V_x^\top\Sig_{xx}\V_x+\V_y^\top\Sig_{yy}\V_y=\I_k$.
We can rewrite the CCA objective as a generalized eigenproblem:
\begin{align}\label{eq:cca1b}
    \argmax{\V_x\in\R^{m\times k},\V_y\in\R^{n\times k}}\tr
    \begin{bmatrix}
        \V_x\\ \V_y
    \end{bmatrix}^\top
    \begin{bmatrix}
         & \Sig_{xy}\\
        \Sig_{xy}^\top&  
    \end{bmatrix}
    \begin{bmatrix}
        \V_x\\ \V_y
    \end{bmatrix},
\end{align}
subject to the whitening constraint
\begin{align}
    \label{eq:cca2b}
    \begin{bmatrix}
        \V_x\\ \V_y
    \end{bmatrix}^\top
    \begin{bmatrix}
        \Sig_{xx} & \\
        & \Sig_{yy}
    \end{bmatrix}
    \begin{bmatrix}
        \V_x\\ \V_y
    \end{bmatrix}
    =\I_k.
\end{align}
The equivalence can be readily verified by expanding the matrix products in Equations~\eqref{eq:cca1b}--\eqref{eq:cca2b}.

Next, we transform this generalized eigenproblem formulation \eqref{eq:cca1b}--\eqref{eq:cca2b} into a standard eigenproblem formulation.
Since the trace of the left hand side of Equation~\eqref{eq:cca2b} is constrained to equal $k$, we can add it to the trace in Equation~\eqref{eq:cca1b}, without affecting the output of the argmax, to arrive at the CCA objective:
\begin{align}
\label{eq:cca3a}
    \argmax{\V_x\in\R^{m\times k},\V_y\in\R^{n\times k}}\tr
    \begin{bmatrix}
        \V_x\\ \V_y
    \end{bmatrix}^\top
    \begin{bmatrix}
        \Sig_{xx} & \Sig_{xy}\\
        \Sig_{xy}^\top & \Sig_{yy}  
    \end{bmatrix}
    \begin{bmatrix}
        \V_x\\ \V_y
    \end{bmatrix}
\end{align}
subject to the whitening constraint in Equation~\eqref{eq:cca2b}.

Our final step is a substitution of variables.
Define the $d\times k$ matrix
\begin{align}
\label{eq:Vxi}
    \V_\xi:=
    \begin{bmatrix}
        \Sig_{xx}^{1/2}\V_x \\ \Sig_{yy}^{1/2}\V_y
    \end{bmatrix}.
\end{align}
After substituting into Equations~\eqref{eq:cca3a} and \eqref{eq:cca2b}, we see that $(\ol\V_x,\ol\V_y)$ is a solution of the CCA objective if and only if $\ol\V_\xi$ is a solution of:
\begin{align}
\label{eq:cca_concat1}
    \argmax{\V_\xi\in\R^{d\times k}}\tr\V_\xi^\top\Sig_{\xi\xi}\V_\xi
    \qquad\text{subject to}\qquad\V_\xi^\top\V_\xi=\I_k,
\end{align}
where we recall that $\Sig_{\xi\xi}$ is the covariance matrix defined by $\Sig_{\xi\xi}:=\frac1T\sum_{t=1}^T\boldsymbol{\xi}_t\boldsymbol{\xi}_t^\top$.
Importantly, Equation~\eqref{eq:cca_concat1} is the variance maximization objective for the standard PCA eigenproblem, which is optimized when the column vectors of $\V_\xi$ form an orthonormal basis spanning the $k$-dimensional principal subspace of the data $\boldsymbol{\xi}_1,\dots,\boldsymbol{\xi}_T$.
Therefore, by Equations~\eqref{eq:xi_apdx} and \eqref{eq:Vxi}, the projection of the vector $\boldsymbol{\xi}_t$ onto its principal subspace is precisely the desired sum of CCSPs: 
    $$\z_t:=\ol\V_x^\top\x_t+\ol\V_y^\top\y_t=\ol\V_\xi^\top\boldsymbol{\xi}_t.$$

\section{Adaptive Bio-CCA with output whitening}

\subsection{Detailed derivation of Algorithm \ref{alg:adaptive}}
\label{apdx:extension}

Recall the min-max objective given in Equation \eqref{eq:minmaxtildeL}:
\begin{align}
    \argmin{\Z\in\R^{k\times T}}\max_{\N\in\R^{k\times T}}\min_{\W_x\in\R^{k\times m}}\min_{\W_y\in\R^{k\times n}}\max_{\bfP\in\R^{k\times k}} \widetilde{L}(\W_x,\W_y,\bfP,\Z,\N),
\end{align}
where
\begin{align*}
    \widetilde{L}(\W_x,\W_y,\bfP,\Z,\N)&:=\frac{1}{T}\tr\left(-2\Z^\top\W_x\X-2\Z^\top\W_y\Y+\alpha \Z^\top\Z\right)\\
    &\qquad+\frac1T\tr\left(2\N^\top \bfP^\top\Z - \N^\top\N\right)\\
    &\qquad+\tr\left(\W_x\Sig_{xx}\W_x^\top+\W_y\Sig_{yy}\W_y^\top-\bfP\bfP^\top\right).
\end{align*}
Since $\widetilde{L}(\W_x,\W_y,\bfP,\Z,\N)$ is strictly convex in $\W_x$, $\W_y$ and $\Z$, and strictly concave in $\bfP$ and $\N$, we can interchange the order of optimization to obtain:
\begin{align}
    \min_{\W_x\in\R^{k\times m}}\min_{\W_y\in\R^{k\times n}}\max_{\bfP\in\R^{k\times k}}\min_{\Z\in\R^{k\times T}}\max_{\N\in\R^{k\times T}}\widetilde{L}(\W_x,\W_y,\bfP,\Z,\N).
\end{align}
The interchange of the maximization with respect to $\N$ and the minimization with respect to $\W_x$ and $\W_y$ is justified by the fact that, for fixed $\Z$ and $\P$, $\widetilde L(\W_x,\W_y,\P,\Z,\N)$ is strictly convex in $(\W_x,\W_y)$ and strictly concave in $\N$.
Similarly, the interchange of the minimization with respect to $\Z$ and the maximization with respect to $\P$ is justified by the fact that, for fixed $\N$, $\W_x$ and $\W_y$, $\widetilde L(\W_x,\W_y,\P,\Z,\N)$ is convex in $\Z$ and strictly concave in $\P$.
In order to derive an online algorithm, we write the objective in terms of time-separable terms:
\begin{align*}
    \widetilde{L}(\W_x,\W_y,\bfP,\Z,\N)=\frac{1}{T}\sum_{t=1}^T\widetilde{l}_t(\W_x,\W_y,\bfP,\z_t,\n_t),
\end{align*}
where
\begin{align*}
    \widetilde{l}_t(\W_x,\W_y,\bfP,\z_t,\n_t)&:=-2\z_t^\top\W_x\x_t-2\z_t^\top\W_y\y_t+\alpha\z_t^\top\z_t+2\n_t^\top\bfP^\top\z_t-\n_t^\top\n_t\\
    &\qquad+\tr\left(\W_x\x_t\x_t^\top\W_x^\top+\W_y\y_t\y_t^\top\W_y^\top-\bfP\bfP^\top\right).
\end{align*}
At each time step $t$, for fixed $\W_x$, $\W_y$ and $\M$, we first simultaneously maximize the objective function $\widetilde{l}_t(\W_x,\W_y,\bfP,\z_t,\n_t)$ over $\n_t$ and minimize over $\z_t$ by running the fast gradient descent-ascent dynamics in Eqs.~\eqref{eq:dzt}--\eqref{eq:dnt} to equilibrium.
Since $\widetilde{l}_t(\W_x,\W_y,\bfP,\z_t,\n_t)$ is convex in $\z_t$ and concave in $\n_t$, these fast dynamics equilibriate at the saddle point where $\ol\z_t=(\bfP\bfP^\top+\alpha\I_k)^{-1}(\a_t+\b_t)$ and $\ol\n_t=\bfP^\top\z_t$.
Then, with $(\ol\n_t,\ol\z_t)$ fixed, we perform a gradient descent-ascent step of the objective function with respect to $(\W_x,\W_y)$ and $\bfP$:
\begin{align*}
    \W_x&\gets\W_x-\frac\eta2\frac{\partial\widetilde{l}_t(\W_x,\W_y,\bfP,\ol\z_t,\ol\n_t)}{\partial\W_x}\\
    \W_x&\gets\W_y-\frac\eta2\frac{\partial\widetilde{l}_t(\W_x,\W_y,\bfP,\ol\z_t,\ol\n_t)}{\partial\W_y}\\
    \bfP&\gets\bfP+\frac{\eta}{2\tau}\frac{\partial\widetilde{l}_t(\W_x,\W_y,\bfP,\ol\z_t,\ol\n_t)}{\partial\bfP}.
\end{align*}
Substituting in with the partial derivatives yields Algorithm~\ref{alg:adaptive}.

\subsection{Comparison with Bio-RRR}
\label{apdx:compare}

In this section, we compare Adaptive Bio-CCA with output whitening (Algorithm \ref{alg:adaptive}) and Bio-RRR (\citep[Algorithm 2]{golkar2020}). 
We first state the Bio-RRR algorithm.\footnote{In \citet{golkar2020} the inputs $\x_t$ are the basal inputs and the inputs $\y_t$ are the apical inputs. Here we switch the inputs to be consistent with Algorithm \ref{alg:adaptive}.}

\begin{algorithm}[ht]
  \caption{Bio-RRR}
  \label{alg:rrr}
\begin{algorithmic}
  \STATE {\bfseries input} data $\{(\x_1,\y_1),\dots,(\x_T,\y_T)\}$; max output-dimension $k$; weight $0\le s\le 1$
  \STATE {\bfseries initialize} weight matrices $\W_x$, $\W_y$, and $\bfP$.
  \FOR{$t=1, 2,\dots,T $}
  \STATE $\a_t\gets\W_x\x_t\quad;\quad\z_t\gets\W_y\y_t\quad;\quad\n_t\gets\bfP^\top\z_t$
  \STATE $\W_x \gets \W_x + \eta_x ( \z_t \x_t^\top- s\a_t \x_t^\top - (1-s)\W_x)$ 
  \STATE $\W_y \gets \W_y + \eta_y ( \a_t-\bfP\n_t )\y_t^\top$ 
  \STATE $\bfP \gets \bfP + \eta_p (\z_t \n_t^\top-\bfP) $
  \ENDFOR
\end{algorithmic}
\end{algorithm}

Bio-RRR implements a supervised learning method for minimizing reconstruction error, with the parameter $0\le s\le 1$ specifying the norm under which the error is measured (see \citep[section 3]{golkar2020} for details).
Importantly, when $s=1$, Algorithm \ref{alg:rrr} implements a supervised version of CCA.
Setting $\alpha=1$ in Algorithm \ref{alg:adaptive} and $s=1$ in Algorithm \ref{alg:rrr}, the algorithms have identical network architectures and synaptic update rules, namely:
\begin{align*}
    \W_x&\gets\W_x+\eta_x(\z_t-\a_t)\x_t^\top\\
    \W_y&\gets\W_y+\eta_y\c_t^a\y_t^\top\\
    \bfP&\gets\bfP+\eta_p(\z_t\n_t^\top-\bfP),
\end{align*}
where we recall that $\c_t^a:=\a_t-\bfP\n_t$.
For this parameter choice, the main difference between the algorithms is that Adaptive Bio-CCA with output whitening is an unsupervised learning algorithm whereas Bio-RRR is a supervised learning algorithm, which is reflected in their outputs $\z_t$: the output of Algorithm \ref{alg:adaptive} is the whitened sum of the basal dendritic currents and the apical calcium plateau potential, i.e., $\z_t=\b_t+\c_t^a$, whereas the output of \cite[Algorithm 2]{golkar2020} is the (whitened) CCSP of the basal inputs, i.e., $\z_t=\b_t$.
In other words, the apical inputs do not directly contribute to the output of the network in \cite{golkar2020}; only indirectly via plasticity in the basal dendrites.
Experimental evidence suggests that apical calcium plateau potentials contribute significantly to the outputs of pyramidal cells, which supports the model derived here. Furthermore, the model in this work allows one to adjust the parameter $\alpha$ to adaptively set the output rank, which is important for analyzing non-stationary input streams.
In Table \ref{tab:comparison} we summarize the differences between the two algorithms.

\begin{table}[ht]
    \centering
    \begin{tabular}{|c|c|c|}\hline
    & Adaptive Bio-CCA & Bio-RRR \\ \hline\hline
    unsupervised/supervised & unsupervised & supervised \\ \hline
    whitened outputs & $\surd$ & $\surd$ \\ \hline
    adaptive output rank & $\surd$ & $\times$ \\ \hline
    \end{tabular}
    \caption{Comparison of Adaptive Bio-CCA with output whitening and Bio-RRR.}
    \label{tab:comparison}
\end{table}

\subsection{Decoupling the interneuron synapses}\label{apdx:decouple}

The neural network for Adaptive Bio-CCA with output whitening derived in Section \ref{sec:extensions} requires the pyramidal neuron-to-interneuron synaptic weight matrix~$\bfP^\top$ to be the the transpose of the interneuron-to-pyramidal neuron synaptic weight matrix~$\bfP$. Enforcing this symmetry via a centralized mechanism is not biologically plausible. Rather, following \citep[Appendix D]{golkar2020}, we show that the symmetry between these two sets of weights naturally follows from the local learning rules.

To begin, we replace the pyramidal neuron-to-interneuron weight matrix~$\bfP^\top$ by a new weight matrix $\bf R$ with Hebbian learning rules:
\begin{align}\label{eq:Rupdate}
    \bf R \gets&\bf R+\frac{\eta}{\tau}(\n_t\z_t^\top-\bf R).
\end{align}
Let $\bfP_0$ and ${\bf R}_0$ denote the initial values of $\bfP$ and $\bf R$.
Then, in view of the updates rule for $\bfP$ in and $\bf R$ in Algorithm \ref{alg:adaptive} and Equation \eqref{eq:Rupdate}, respectively, the difference $\bfP^\top - \bf R$ after $T$ updates is given by
\begin{equation*}
\bfP^\top - {\bf R} = \left(1-\frac\eta\tau\right)^T (\bfP_0^\top-\bf R_0).
\end{equation*}
In particular, the difference decays exponentially (recall that $\eta<\tau$ by assumption).

\section{Numerics}

\subsection{Experimental details}
\label{apdx:details}

\paragraph{Bio-CCA:} We implemented Algorithm~\ref{alg:online}.
We initialized $\W_x$ (rexp.\ $\W_y$) to be a random matrix with i.i.d.\ mean-zero normal entries with variance $1/m$ (resp.\ $1/n$).
We initialized $\M$ to be the indentity matrix $\I_k$.
We used the time-dependent learning rate of the form $\eta_t=\eta_0/(1+\gamma t)$.
To find the optimal hyperparameters, we performed a grid search over $\eta_0\in\{10^{-2},10^{-3},10^{-4}\}$, $\gamma\in\{10^{-3},10^{-4},10^{-5}\}$ and $\tau\in\{1,0.5,0.1\}$.
The best performing parameters are reported in Table \ref{tab:hyper}. 
Finally, to ensure the reported basis vectors satisfy the orthonormality constraints \eqref{eq:cca2}, we report the following normalized basis vectors:
\begin{align}
    \label{eq:Vxnorm}
    \V_x^\top&:=\left\{\M^{-1}(\W_x\Sig_{xx}\W_x^\top+\W_y\Sig_{yy}\W_y^\top)\M^{-1}\right\}^{-1/2}\M^{-1}\W_x,\\
    \label{eq:Vynorm}
    \V_y^\top&:=\left\{\M^{-1}(\W_x\Sig_{xx}\W_x^\top+\W_y\Sig_{yy}\W_y^\top)\M^{-1}\right\}^{-1/2}\M^{-1}\W_y.
\end{align}

\begin{table}[ht]
    \centering
    \begin{tabular}{|c|c|c|}\hline
        & synthetic & \texttt{Mediamill} \\ \hline\hline
        Bio-CCA $(\eta_0,\gamma,\tau)$ & $(10^{-3},10^{-4},0.1)$ & $(10^{-2},10^{-4},0.1)$ \\ \hline
        Gen-Oja $(\beta_0,\gamma)$ & $(1,10^{-2})$ & $(10^{-2},10^{-3})$ \\ \hline
        Asym-NN $(\eta_0,\alpha)$ & $(2\times10^{-4},5\times10^{-6})$ & $(2\times10^{-3},5\times10^{-6})$ \\ \hline
        Bio-RRR $(\eta_0,\gamma,\mu)$ & $(10^{-3},10^{-4},10)$ & $(10^{-2},10^{-5},10)$ \\ \hline
        Adaptive Bio-CCA $(\eta_0,\gamma,\tau)$ & $(10^{-3},10^{-4},0.1)$ & $(10^{-2},10^{-4},0.5)$ \\ \hline
    \end{tabular}
    \caption{Optimal time-dependent learning rates.}
    \label{tab:hyper}
\end{table}

\paragraph{MSG-CCA:} We implemented the online algorithm stated in \citep{arora2017stochastic}. 
MSG-CCA requires a training set to estimate the covariance matrices $\Sig_{xx}$ and $\Sig_{yy}$.
We provided the algorithm with 1000 samples to initially estimate the covariance matrix.
Following \citep{arora2017stochastic}, we use the time-dependent learning rate $\eta_t=0.1/\sqrt{t}$.

\paragraph{Gen-Oja:} We implemented the online algorithm stated in \citep{bhatia2018gen}.
The algorithm includes 2 learning rates: $\alpha_t$ and $\beta_t$. As stated in \citep{bhatia2018gen}, the Gen-Oja's performance is robust to changes in the learning rate $\alpha_t$, but sensitive to changes in the learning rate $\beta_t$. Following \citep{bhatia2018gen}, we set $\alpha_t$ to be constant and equal to $1/R^2$ where $R^2:=\tr(\Sig_{xx})+\tr(\Sig_{yy})$. To optimize over $\beta_t$, we used a time-dependent learning rate of the form $\beta_t=\beta_0/(1+\gamma t)$ and performed a grid search over $\beta_0\in\{1,10^{-1},10^{-2}\}$ and $\gamma\in\{10^{-1},10^{-2},10^{-3}\}$. The best performing parameters are reported in Table \ref{tab:hyper}.

\paragraph{Asymmetric CCA Network:} We implemented the online multi-channel CCA algorithm derived in \cite{Zhao2020}. Following \cite{Zhao2020}, we use the linearly decaying learning rate $\eta_t=\eta_0\times\max(1-\alpha t,0.1)$. To optimize the performance of the algorithm we performed a grid search over $\eta_0\in\{2\times10^{-2},10^{-2},2\times10^{-3},10^{-3}\}$ and $\alpha\in\{5\times10^{-5},10^{-5},5\times10^{-6},10^{-6}\}$. The best performing parameters are reported in Table \ref{tab:hyper}.

\paragraph{Bio-RRR:} We implemented the online CCA algorithm derived in \citep{golkar2020} with $s=1$ (see Algorithm \ref{alg:rrr}). The algorithm includes learning rates $\eta_x$, $\eta_y$ and $\eta_p$. Following \citep{golkar2020}, we set $\eta_y=\eta_t$ and $\eta_x=\eta_p=\eta_y/\mu$ and use the time-dependent learning rate of the form $\eta_t=\eta_0/(1+\gamma t)$. We performed a grid search over $\eta_0\in\{10^{-2},10^{-3},10^{-4}\}$, $\gamma\in\{10^{-3},10^{-4},10^{-5}\}$ and $\mu\in\{1,10,100\}$ and list best performing parameters in Table \ref{tab:hyper}.

\paragraph{Adaptive Bio-CCA with output whitening:} We implemented Algorithm~\ref{alg:adaptive}.
We initialized $\W_x$, $\W_y$ and $\P$ to be random matrices with i.i.d.\ standard normal entries.
To find the optimal hyperparameters, we performed a grid search over $\eta_0\in\{10^{-2},10^{-3},10^{-4}\}$, $\gamma\in\{10^{-3},10^{-4},10^{-5}\}$ and $\tau\in\{1,0.5,0.1\}$.
The best performing parameters are reported in Table \ref{tab:hyper}. 

\subsection{Orthonormality constraints}
\label{apdx:ortho}

\begin{figure}
    \centering
    \includegraphics[width=\textwidth]{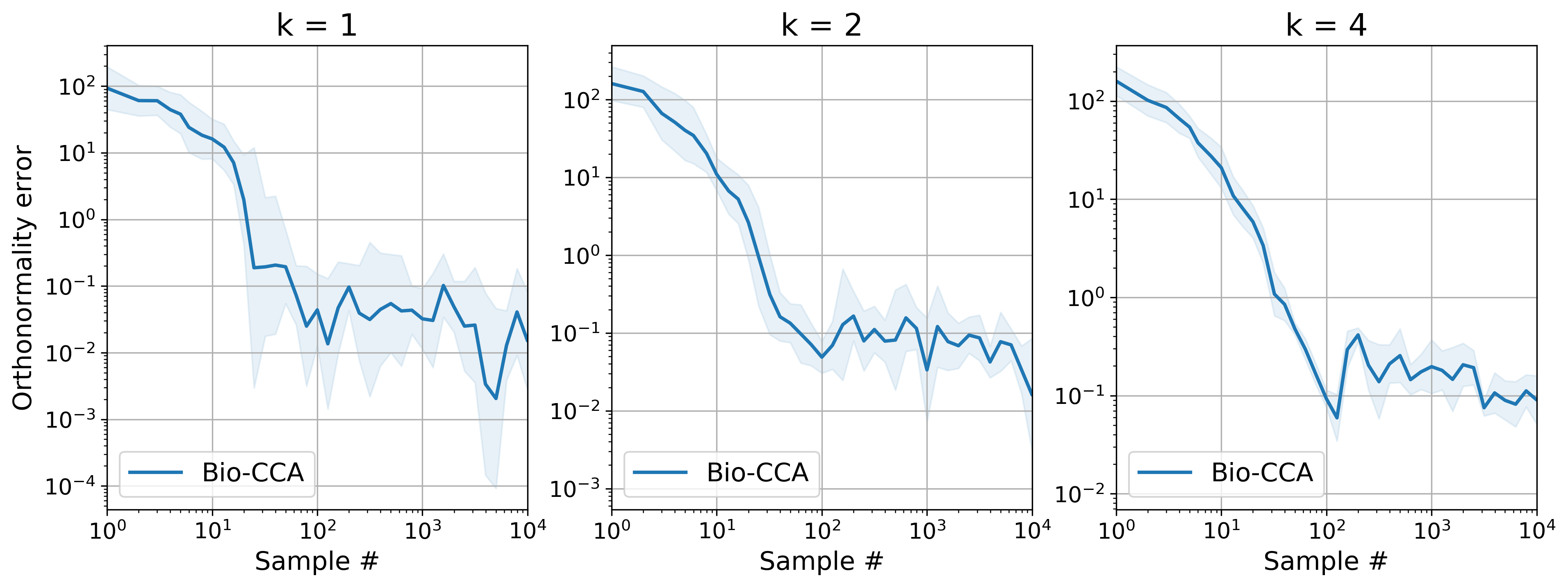}
    \caption{The deviation of the Bio-CCA solution from the CCA orthonormality constraint, in terms of the orthonormality error defined in Equation~\eqref{eq:ortho}, on the synthetic dataset.} 
\label{fig:ortho1}
\end{figure}

\begin{figure}
    \centering
    \includegraphics[width=\textwidth]{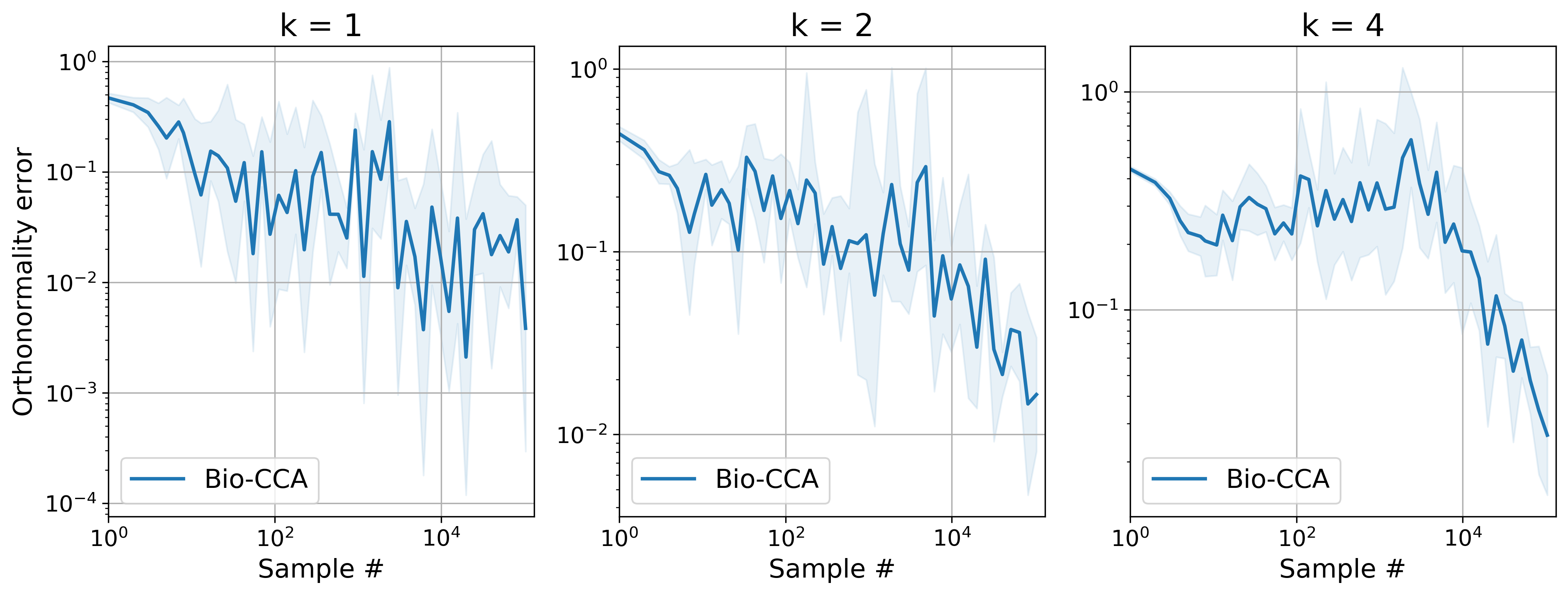}
    \caption{The deviation of the Bio-CCA solution from the CCA orthonormality constraint, in terms of the orthonormality error defined in Equation~\eqref{eq:ortho}, on the dataset \texttt{Mediamill}.} 
\label{fig:ortho2}
\end{figure}

\paragraph{Bio-CCA.} To verify that the Bio-CCA basis vectors asymptotically satisfy the orthonormality constraint \eqref{eq:cca2}, we use the following orthonormality error function:
\begin{align}
    \label{eq:ortho}
    \text{Orthonormality Error}(t):=\frac{\|\M_t^{-1}(\W_{x,t}\Sig_{xx}\W_{x,t}^\top+\W_{y,t}\Sig_{yy}\W_{y,t}^\top)\M_t^{-1}-\I_k\|_F^2}{k}.
\end{align}
In Figure~\ref{fig:ortho1} (resp.\ Figure \ref{fig:ortho2}) we demonstrate that Bio-CCA asymptotically satisfies the CCA orthonormality constraints \eqref{eq:cca2} on the synthetic dataset (resp.\ \texttt{Mediamill}).

\begin{figure}
    \centering
    \includegraphics[width=\textwidth]{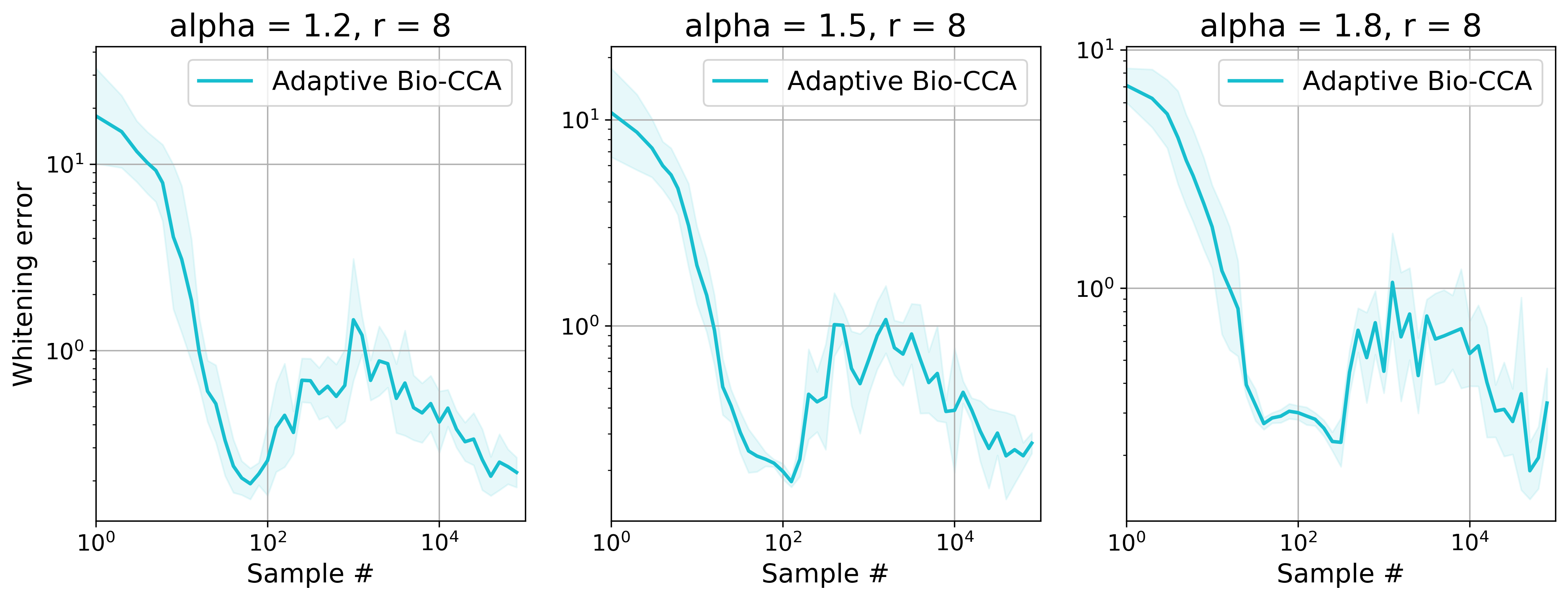}
    \caption{The deviation of the Bio-CCA solution from the CCA orthonormality constraint, in terms of the orthonormality error defined in Equation~\eqref{eq:ortho}, on the synthetic dataset.} 
\label{fig:white1}
\end{figure}

\begin{figure}
    \centering
    \includegraphics[width=\textwidth]{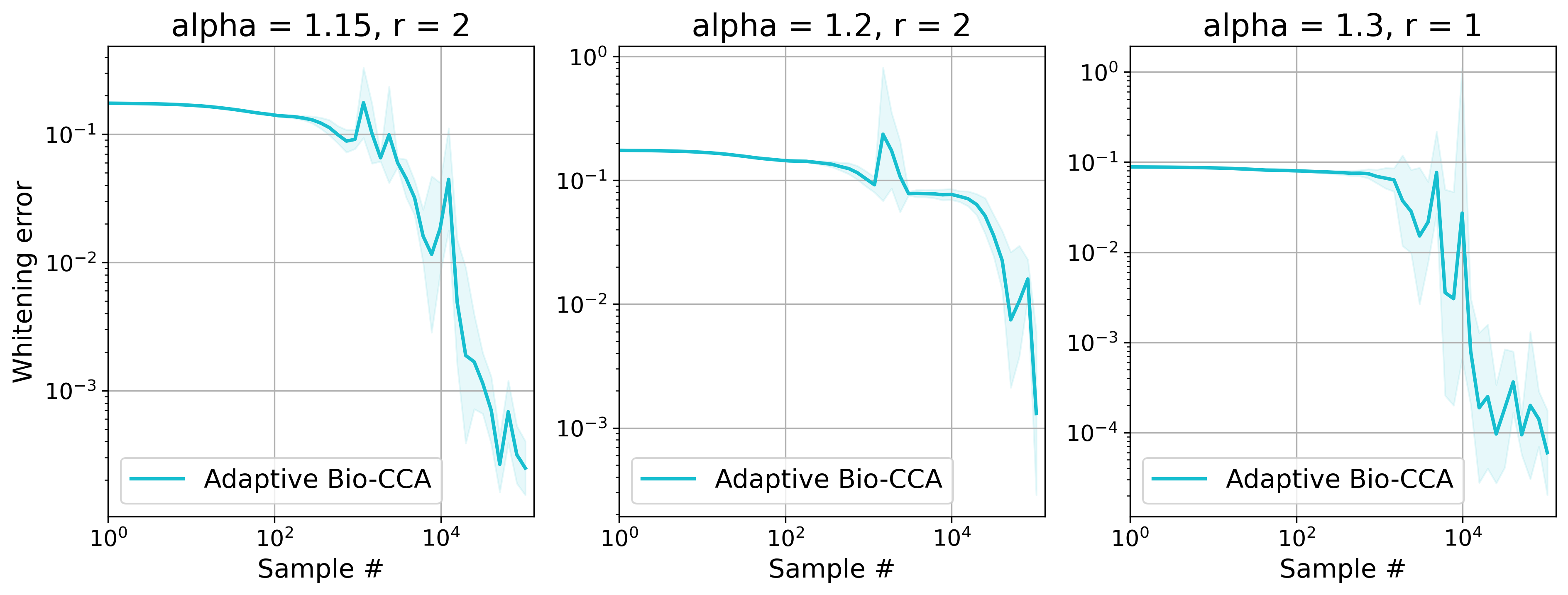}
    \caption{The deviation of the Adaptive Bio-CCA with output whitening solution from the whitening constraint, in terms of the whitening error defined in Equation~\eqref{eq:whitening_error}, on the dataset \texttt{Mediamill}.} 
\label{fig:white2}
\end{figure}

\paragraph{Adaptive Bio-CCA with output whitening.} To verify that the top $r$ eigenvalues of the output covariance asymptotically approach 1, we let $\lambda_1(t)\ge\cdots\ge\lambda_k(t)$ denote the ordered eigenvalues of the matrix
\begin{align*}
    \Sig_{zz}(t):=\V_{x,t}^\top\Sig_{xx}\V_{x,t}+\V_{x,t}^\top\Sig_{xy}\V_{y,t}+\V_{y,t}^\top\Sig_{xy}^\top\V_{x,t}+\V_{y,t}^\top\Sig_{yy}\V_{y,t},
\end{align*}
and define the whitening error by
\begin{align}\label{eq:whitening_error}
    \text{Whitening Error}(t)&:=\frac{\sum_{i=1}^{r}|\lambda_i(t)-1|^2+\sum_{i=r+1}^k|\lambda_i(t)|^2}{k}.
\end{align}
In Figure~\ref{fig:white1} (resp.\ Figure~\ref{fig:white2}) we demonstrate that Bio-CCA asymptotically satisfies the CCA orthonormality constraints \eqref{eq:cca2} on the synthetic dataset (resp.\ \texttt{Mediamill}).

\section{On convergence of the CCA algorithms}
\label{apdx:gda}

To interpret our offline and online CCA algorithms (Algorithms \ref{alg:offline} \& \ref{alg:online}) mathematically, we first make the following observation.
Both algorithms optimize the min-max objective \eqref{eq:minmax}, which includes optimization over the neural outputs $\Z$.
In this way, the neural activities can be viewed as optimization steps, which is useful for a biological interpretation of the algorithms.
However, since we assume a separation of time-scales in which the neural outputs equilibrate at their optimal values before the synaptic weight matrices are updated, the neural dynamics are superfluous when analyzing the algorithms from a mathematical perspective.
Therefore, we set $\Z$ equal to its equilibrium value $\ol\Z=\M^{-1}(\W_x\X+\W_y\Y)$ in the cost function $L(\W_x,\W_y,\M,\Z)$ to obtain a min-max problem in terms of the synaptic weights:
\begin{align}
    \label{eq:minmaxWM}
    \min_{\W\in\R^{k\times d}}\max_{\M\in\mathcal{S}_{++}^k}F(\W,\M),
\end{align}
where $\W:=[\W_x\;\W_y]$ is the matrix of concatenated weights and $F(\W,\M):=L(\W_x,\W_y,\M,\ol\Z)$.
After substitution, we see that $F:\R^{k\times d}\times\mathcal{S}_{++}^k\to\R$ is the nonconvex-concave function
    $$F(\W,\M)=\tr\left(-\M^{-1}\W\A\W^\top+\W\B\W^\top-\frac12\M^2\right),$$
with partial derivatives
\begin{align*}
    -\frac{\partial F(\W,\M)}{\partial \W}&=2\M^{-1}\W\A-2\W\B\\
    \frac{\partial F(\W,\M)}{\partial \M}&=\M^{-1}\W\A\W^\top\M^{-1}-\M.
\end{align*}
where we have defined
\begin{align}
\label{eq:CD}
    \A:=\begin{bmatrix}\Sig_{xx} & \Sig_{xy} \\ \Sig_{xy}^\top & \Sig_{yy} \end{bmatrix},&&\B:=\begin{bmatrix}\Sig_{xx} & \\ & \Sig_{yy}\end{bmatrix}.
\end{align}
Similar objectives, with different values of $\A$ and $\B$, arise in the analysis of online principal subspace projection \cite{pehlevan2018similarity} and slow feature analysis \cite{lipshutz2020biologically} algorithms.

The synaptic updates in both our offline and online algorithms can be viewed as (stochastic) gradient descent-ascent algorithms for solving the noncovex-concave min-max problem \eqref{eq:minmaxWM}.
To make the comparison with our offline algorithm, we substitute the optimal value $\ol\Z$ into the synaptic weight updates in Algorithm~\ref{alg:offline} to obtain:
\begin{align}\label{eq:offlineWM1}
    \W&\gets\W+2\eta(\M^{-1}\W\A-\W\B)\\
    \label{eq:offlineWM2}
    \M&\gets\M+\frac{\eta}{\vareps}(\M^{-1}\W\A\W^\top\M^{-1}-\M),
\end{align}
Comparing these updates to the partial derivatives of $F$, we see that Offline-CCA is naturally interpreted as a gradient descent-ascent algorithm for solving the min-max problem \eqref{eq:minmaxWM}, with descent step size $\eta$ and ascent step size $\frac\eta\vareps$.
Similarly, to make the comparison with our online algorithm, we substitute the explicit expression for the equilibrium value $\ol\z_t=\M^{-1}(\a_t+\b_t)$, where $\a_t=\W_x\x_t$ and $\b_t=\W_y\y_t$, into the synaptic weight updates in Algorithm~\ref{alg:online} to rewrite the updates:
\begin{align*}
    \W&\gets\W+2\eta(\M^{-1}\W\A_t-\W\B_t)\\
    \M&\gets\M+\frac{\eta}{\vareps}(\M^{-1}\W\A_t\W^\top\M^{-1}-\M),
\end{align*}
where
\begin{align*}
    \A_t:=\begin{bmatrix}\x_t\x_t^\top & \x_t\y_t^\top \\ \y_t\x_t^\top & \y_t\y_t^\top\end{bmatrix},&&\B_t:=\begin{bmatrix}\x_t\x_t^\top & \\ & \y_t\y_t^\top\end{bmatrix}.
\end{align*}
Comparing these updates to the partial derivatives of $F$, we see that our online algorithm is naturally interpreted as a stochastic gradient descent-ascent algorithm for solving the min-max problem \eqref{eq:minmaxWM}, using the time $t$ rank-1 approximations $\A_t$ and $\B_t$ in place of $\A$ and $\B$, respectively.

Establishing theoretical guarantees for solving nonconvex-concave min-max problems of the form \eqref{eq:minmaxWM} via stochastic gradient descent-ascent is an active area of research \citep{razaviyayn2020nonconvex}.
\citet{borkar1997stochastic,borkar2009stochastic} proved asymptotic convergence to the solution of the min-max problem for a two time-scale stochastic gradient descent-ascent algorithm, where the ratio between the learning rates for the minimization step and the maximization step, $\vareps$, depends on the iteration and converges to zero in the limit as the iteration number approaches infinity.
\citet{lin2019gradient} established convergence rate guarantees for a stochastic gradient descent-ascent algorithm to an equilibrium point (not necessarily a solution of the min-max problem).
Both these results, however, impose assumptions that do not hold in our setting: the partial derivatives of $F$ are Lipschitz continuous and $\M$ is restricted to a bounded convex set.
Therefore, establishing global stability with convergence rate guarantees for our offline and online CCA algorithms requires new mathematical techniques that are beyond the scope of this work.

Even proving local convergence properties is non-trivial.
In the special case that $\B=\I_d$, \citet{pehlevan2018similarity} carefully analyzed the continuous dynamical system obtained by formally taking the step size $\eta$ to zero in Equations \eqref{eq:offlineWM1}--\eqref{eq:offlineWM2}.
They computed an explicit value $\tau_0\ge1/2$, in terms the eigenvalues of $\A$ such that if $\tau\le\tau_0$, then solutions of the min-max problem \eqref{eq:minmaxWM} are the only linearly stable fixed points of the continuous dynamics.
The case that $\B\ne\I_d$ is more complicated, and the approach in \citep{pehlevan2018similarity} is not readily extended.
In ongoing work, we take a step towards understanding the asymptotics of our algorithms by analyzing local stability properties for a general class of gradient descent-ascent algorithms, which includes Offline-CCA and Bio-CCA as special cases.

\bibliography{CCA.bib}
\bibliographystyle{plainnat}

\end{document}